\documentclass[11pt,a4paper]{article}
\pdfoutput=1
\usepackage{jheppub}
\usepackage{amsmath,amsfonts,amssymb,bbm}
\usepackage{hyperref, csquotes}
\usepackage{graphicx, color}
\usepackage{tikz,pgf}

\usepackage{cancel}

\tikzstyle{v} =  [circle, draw=black, line width=.2pt, fill=black, inner sep=0pt, minimum size=1.5mm]
\tikzstyle{wv} = [circle, inner sep=0.1pt, draw=jred, minimum size=2mm]
\tikzstyle{bv} = [circle, inner sep=0.1pt, fill=jred, minimum size=2mm]
\tikzstyle{e} =	[draw=jred,line width=1pt]
\tikzstyle{eb}= [draw=jgreen,line width=1pt]
\usepackage{physics,tensor}

\renewcommand{\O}{\textrm{O}}
\newcommand{\U}{\textrm{U}}
\newcommand{\R}{ {\mathbb{R}} }
\newcommand{\C}{ {\mathbb{C}}}
\newcommand{\Z}{{\mathbb{Z}}}

\renewcommand{\eqref}[1]{Eq.\,(\ref{#1})}
\newcommand{\beq}{ \begin{equation} }
\newcommand{\eeq}{ \end{equation} }
\newcommand{\bea}{\begin{eqnarray}}
\newcommand{\eea}{\end{eqnarray}}

\newcommand{\inter}{{\rm inter}}

\definecolor{jred}{rgb}{0.8,0,0}
\definecolor{jgreen}{rgb}{0,0.7,0}
\definecolor{jblue}{rgb}{0,0,0.8}

\renewcommand\[{\begin{equation}}
\renewcommand\]{\end{equation}}
\newcommand{\boxd}[1]{\boxed{\phantom{\Biggl(}#1\phantom{\Biggl)}}}
\newcommand{\STr}{\mathrm{STr}}
\def\extd{\mathrm {d}}
\newcommand{\id}{\mathbb I}

\newcommand{\ld}{{d}}
\newcommand{\rk}{r}         		
\newcommand{\ks}{{2}}          
\newcommand{\js}{{2\zeta}}          
\newcommand{\ps}{{2}}          
\newcommand{\cs}{V_{G}} 
\newcommand{\kap}{\bar{\kappa}}
\newcommand{\lv}{a_{\R}}    

\newcommand{\wfr}{Z_k}
\newcommand{\wfp}{Z_{p,k}}
\newcommand{\wfj}{Z_{\rep,k}}

\newcommand{\m}{\mu_k}
\newcommand{\mr}{\tilde{\mu}_k}

\newcommand{\mc}{\bar{\mu}_*}
\newcommand{\rhr}{\tilde{\rho}}
\newcommand{\cnk}[1]{\lambda^c_{#1,k}}
\newcommand{\cn}[1]{\lambda_{#1,k}}   
\newcommand{\cnr}[1]{\tilde{\lambda}_{#1,k}}

\newcommand{\reg}{{\mathcal R}_k}
\renewcommand{\and}{\eta_k}
\newcommand{\efd}{{d_{\textrm{eff}}}}
\newcommand{\tfd}{d_{\rk}}

\newcommand{\crd}{d_{\textrm{crit}}}

\newcommand{\gfc}{W}                
\newcommand{\gfi}{\Gamma}           
\newcommand{\gfk}{\Gamma_k}         
\newcommand{\kin}{\mathcal{K}}                
\newcommand{\pr}{P_{\textsc{r}}}
\newcommand{\nmax}{n}               
\newcommand{\ak}{{\tilde{k}}}  

\newcommand{\cct}{}
\newcommand{\ccu}{2}
\newcommand{\gp}{}
\newcommand{\gf}{\Phi}       		
\newcommand{\gfb}{\bar\Phi}

\newcommand{\cgf}{\chi}       		
\newcommand{\cgfb}{\bar\chi}
\newcommand{\vx}{\pmb{x}}
\newcommand{\vy}{\pmb{y}}
\newcommand{\vp}{\pmb{p}}
\newcommand{\vq}{\pmb{q}}
\newcommand{\vg}{\pmb{g}}
\newcommand{\vh}{\pmb{h}}
\newcommand{\dg}[1]{\delta(g_{#1},h_{#1})}
\newcommand{\mop}{(\gfb\cdot_{\hat{c}}\gf)}
\newcommand{\nop}{(\gfb\cdot_{{c}}\gf)}
\newcommand{\rep}{j}

\newcommand{\cas}[1]{C^{(\zeta)}_{#1}}
\newcommand{\vrep}{{\pmb{\rep}}}

\newcommand{\vol}[1]{v_{#1}}
\newcommand{\vks}[1]{v_{#1}^{(\zeta)}}

\newcommand{\kw}[1]{F^{(#1)}}
\newcommand{\kf}[1]{I^{(#1)}}
\newcommand{\ku}[1]{I_1^{(#1)}}
\newcommand{\kp}[1]{I_{p^2}^{(#1)}}
\newcommand{\kj}[1]{I_{\rep^\js}^{(#1)}}
\newcommand{\fr}{F}

\newcommand{\pot}{U_k}
\newcommand{\self}[1]{M_k^{(#1)}}

\newcommand{\bca}[1]{\beta^{\textsc{v1}}_{#1}}
\newcommand{\bcb}[1]{\beta^{\textsc{v2}}_{#1}}
\newcommand{\bfn}[1]{\beta^{\textsc{cm}}_{#1}}

\renewcommand{\aa}{a}
\newcommand{\bb}{b}

\begin{document}

\title{QFT with Tensorial and Local Degrees of Freedom: 
Phase Structure from Functional Renormalization}
\author[a,b]{Joseph Ben Geloun,}
\author[c,d]{Andreas G. A. Pithis,}
\author[e]{Johannes Th{\"u}rigen,}

\emailAdd{bengeloun@lipn.univ-paris13.fr}
\emailAdd{andreas.pithis@physik.lmu.de}
\emailAdd{johannes.thuerigen@uni-muenster.de}

\affiliation[a]{Laboratoire d’Informatique de Paris Nord UMR CNRS 7030
Universit\'{e} Paris 13,\\ 99, avenue J.-B. Clement, 93430 Villetaneuse, France}
\affiliation[b]{International Chair in Mathematical Physics and Applications
ICMPA–UNESCO Chair,\\ 072 B.P. 50 Cotonou, Benin}
\affiliation[c]{Arnold Sommerfeld Center for Theoretical Physics, Ludwig-Maximilians-Universit\"at München,\\ Theresienstrasse 37, 80333 M\"unchen, Germany, EU}
\affiliation[d]{Munich Center for Quantum Science and Technology (MCQST),\\ Schellingstr. 4, 80799 M\"unchen, Germany, EU}
\affiliation[e]{Mathematisches Institut der Westf\"alischen Wilhelms-Universit\"at M\"unster,\\ Einsteinstr. 62, 48149 M\"unster, Germany, EU}

\date{\today}
\begin{abstract}{
Field theories with combinatorial non-local interactions such as tensor invariants are interesting candidates for describing a phase transition from discrete quantum-gravitational to continuum geometry.
In the so-called cyclic-melonic potential approximation of a tensorial field theory on the $\rk$-dimensional torus 
it was recently shown using functional renormalization group techniques that no such phase transition to a condensate phase with a tentative continuum geometric interpretation is possible. 
Here, keeping the same approximation, we show how to overcome this limitation amending the theory by local degrees freedom on $\R^\ld$. 
We find that the
effective $\rk-1$ dimensions of the torus part dynamically vanish along the renormalization group flow while the $\ld$ local dimensions persist up to small momentum scales.
Consequently, for $\ld>2$ one can find a phase structure allowing also for phase transitions.
}
\end{abstract}

\setcounter{tocdepth}{2}

\maketitle

\section{Introduction}

Field theories with tensorial interactions provide a promising framework for quantum field theory beyond local, standard-model type quantum field theory.
They are generalizations of 
Kontsevich-type field theories \cite{Kontsevich1992} and 
non-commutative field theories \cite{Grosse:2004yu, Grosse:2006hh, Rivasseau:2007ab}
from matrix fields to tensor fields of order $\rk>2$.
Like the former they come with additional structure in their perturbative expansion provided by a large-$N$ expansion.
This leads to an improved control on the definition of such field theories and evaluation of observables, raising the expectation to provide even solvable models \cite{Bonzom:2012hw}. 
Furthermore, they also provide models of asymptotically free field theory~\cite{BenGeloun:2013vwi,BenGeloun:2012pu,Rivasseau:2015im}.
In addition, as their diagrams are bijective to $\rk$-dimensional discrete manifolds~\cite{Gurau:2009tw,Gurau:2011xp,GurauBook}, they set the stage to consider random geometry in higher dimensions, and in particular quantum gravity when adding geometric degrees of freedom as in tensorial group field theory (TGFT)~\cite{Freidel:2005jy,Oriti:2012wt,Carrozza:2013uq}.

For all these reasons it is a crucial challenge to understand the phase diagram of tensorial field theories beyond perturbation theory.
To this end, functional renormalization group (FRG) techniques have been applied to tensor models \cite{Eichhorn:2013isa, Eichhorn:2014xaa, Eichhorn:2017xhy, Eichhorn:2019hsa, Eichhorn:2020mte, Eichhorn:2020sla, Castro:2020dzt} and field theories with tensorial interactions \cite{Benedetti:2015et, Benedetti:2015yaa, Lahoche:2016xiq, BenGeloun:2015ej, BenGeloun:2016kw, Carrozza:2016tih,Carrozza:2017vkz,BenGeloun:2018ekd,Lahoche:2019orv,Baloitcha:2020idd,Pithis:2020sxm,Pithis:2020kio} up to order $\phi^8$ in the vertex expansion.
Dominance of the subclass of so called ``cyclic-melonic'' interactions allows to apply a local-potential approximation (LPA) \cite{Carrozza:2016tih} in which the phase diagram of rank-$\rk$ tensorial fields turns out to be closely related to that of $\rk-1$ dimensional $\O(N)$-invariant scalar field theory in the large-$N$ limit \cite{Pithis:2020sxm,Pithis:2020kio}. 
In this sense, one can call $\tfd:=\rk-1$ the \emph{effective dimension} of the cyclic-melonic field theory, $\efd=\tfd$.
In particular, there is a Wilson-Fisher type fixed point for $2<\efd<4$.
This is, however, only true upon removing a finite-volume regularization $V\to\infty$ thereby decompactifying the domain $\U(1)\to\R$.
On a compact domain with finite volume~$V$, the common wisdom that a field theory becomes effectively zero-dimensional at small momentum scales (IR) and no transition to a condensate phase is possible \cite{Zinn-Justin:2002ecy,strocchi2005symmetry,Benedetti:2014qsa} applies also to tensorial fields, i.e.~the effective dimension becomes $\efd\to0$ \cite{Pithis:2020sxm,Pithis:2020kio}. Assuming that such non-perturbative vacua have a tentative continuum geometric interpretation, it is important to understand under which conditions such phase transitions can be realized.
Since most related models of quantum gravity considered in the literature restrict to a compact Lie group $G$ \cite{Perez:2012wv,Oriti:2012wt,Carrozza:2013uq}, this raises the question whether a phase transition to continuum geometry is possible at all.

Two ways out of this have been suggested and shown to indeed solve it in a Gaussian approximation, that is, Landau-Ginzburg mean-field theory \cite{Pithis:2018eaq,Pithis:2019mlv,Marchetti:2021wp,Marchetti:2022igl,Marchetti:2022nrf}:
First, the introduction of $\ld$-dimensional scalar degrees of freedom with local, point-like interactions simply adds to the effective dimension $\efd  = \tfd +\ld$ \cite{Marchetti:2021wp}. Beyond leading us to the realm of theories
reminiscent of the Sachdev-Ye-Kitaev (SYK) model~\cite{Rosenhaus:2018dtp} and SYK-type models~\cite{Delporte:2018iyf,Delporte:2020rce}, such degrees of freedom have a natural interpretation of coupling scalar matter as reference frame fields to geometric degrees of freedom in TGFT \cite{Li:2017uao,Oriti:2016qtz,Gielen:2018fqv}.
Then, even if the dimension of the tensorial part becomes zero in the IR, phase transitions are possible with sufficiently large~$\ld$.
Second, there are models of quantum gravity based on $G=\textrm{SL}(2,\R)$ or $G=\textrm{SL}(2,\C)$ related to the Lorentz group; due to the hyperbolic geometry of such group one finds an effective dimension which becomes $\efd\to\infty$ in the IR such that these models become Gaussian and have a mean-field phase transition irrespective of the rank $\rk$ of the tensor field \cite{Marchetti:2022igl,Marchetti:2022nrf}.
In this work, we aim to examine whether and how the first result extends beyond Gaussian approximation.
The equally interesting second case will be left for future work.

To test the effect of additional local degrees of freedom in a tensorial field theory, we consider 
the functional renormalization group of a complex tensor field of rank $\rk$ augmented with point-like interacting degrees of freedom on $\R^\ld$ in the cyclic-melonic potential approximation (LPA).
This is a new type of field theory in that the domain of a single field splits into two distinguished parts, the local and the tensorial variables. 
The free, kinetic theory can have different properties in the two parts
leading to two wave-function renormalization parameters.
In particular, we consider a distinguished scaling with power~$\js$ of its spectrum in the tensorial part.

We find that the local degrees of freedom on $\R^\ld$ simply add to the effective dimension of the tensorial part, $\efd = \tfd/\zeta + \ld$, as 
expected from the previous results in the Gaussian approximation \cite{Marchetti:2021wp}.
This is a direct consequence of the scaling of couplings needed to turn the FRG flow equations into dimensionless equations upon rescaling. 
Accordingly, the critical dimension 
$\efd=4$ occurs for various combinations of tensor rank $\rk$ and local dimension~$\ld$.
In particular, if the tensorial degrees of freedom are compact, only their contribution $\tfd$ to the effective dimension $\efd$ vanishes such that $\efd\to\ld$ towards the IR. This leaves room for an interesting phase structure also in this case. 

Beyond the Gaussian regime we find a Wilson-Fisher type point for $\ks<\efd<4$, 
similar to the purely tensorial theory \cite{Pithis:2020sxm,Pithis:2020kio}. 
The FRG equations differ from those of large-$N$ vector field theory by a relative factor $\rk$ leading to a quantitatively new non-Gaussian fixed point with the qualities of the Wilson-Fisher fixed point. 
In contrast to the purely tensorial case \cite{Pithis:2020sxm,Pithis:2020kio} where $\efd=(\rk-1)/\zeta$, this factor $\rk$ can now have different values for given $\efd$, e.g.~$\rk=2,3,4$ for $\efd=3$ and $\zeta=1$, which leads to quantitatively new non-Gaussian fixed points here.

There are hints that there are more interesting non-Gaussian fixed point in the LPA$'$ even beyond the critical dimension $\efd>4$. 
In the LPA$'$ one includes the flow of the wave-function renormalization and thereby a dynamic anomalous dimension.
From this perspective the extra freedom of two relevant theory parameters $\rk$ and $\ld$ would become even more interesting. 
However, the occurrence of two wave-function renormalizations $\wfp\ne\wfj$ in one single kinetic term leads to new technical challenges which we will address in future work.

\

The structure of the paper is the following:
In Sec.~2 we introduce field theory of both tensorial and local degrees of freedom and show how the FRG method applies in this framework. 
Sec.~3 then carries out the calculations leading to the FRG equation in the local potential approximation, in particular for the cyclic-melonic regime of the theory, and derives explicit beta functions.
In Sec.~4 we discuss then the resulting phase structure before we conclude in Sec.~5. The Appendices complement the main body of this work with
more detailed calculations needed for the computation of the FRG equations.

\section{FRG for theories with local and non-local degrees of freedom}

\subsection{Fields with local and non-local degrees of freedom}

In this section we set up our notation. In particular, the fields, action and flow equation are presented. 

We consider
real- or complex-valued fields $\Phi: \mathbb{R}^\ld \times G^{\times \rk} \to \mathbb{K}=\R, 
\C$,  of $\ld+\rk$ arguments $\vx\in \mathbb{R}^\ld$ and $\vg\in G^{\times \rk}$ where $G$ is a compact Lie group which, later on, we fix to $\text{U}(1)$. These are chosen square-integrable $L^2(\mathbb{R}^\ld 
 \times G^{\times \rk})$ with respect to the inner product
(with adapted consideration for real fields)
\begin{equation}
    (\gf,\gf')= \int_{\R^\ld} \extd\vx \int_{G^\rk} \extd\vg~ \bar{\Phi}(\vx,\vg)\Phi'(\vx,\vg)
\end{equation}
with Lebesgue measure $\extd\vx$ on $\R^\ld$ and dimensionful Haar measure $\extd\vg = (\text{d} g)^r$ 
on $G^{\times r}$ defined on a single copy of $G$ as 
\begin{equation}
    \int_G \text{d} g= \cs.
\end{equation}
An interesting case consists in 
inspecting 
the large volume limit $\cs\to \infty$ which relates to the decompactification of $G^{\times r}$ sending its radius to infinity.  
When $G^{\times r}=\text{U}(1)^r$, for instance, then 
letting the radius of $S^1 \sim \text{U}(1)$ go to infinity,
leads us
from $G^r$ to $\mathbb{R}^r$. This sometimes could be also understood, in a reverse way, 
starting from $\mathbb{R}^r$, 
we introduce a lattice regularization of that theory, i.e. a theory on the discrete momentum space of 
the compact group $G^r$. 
Letting the radius of $G^r$ 
going to infinity while keeping
the lattice spacing going to zero 
is performing  the thermodynamic limit of that regularized theory
\cite{BenGeloun:2015ej, BenGeloun:2016kw}. 

We expand the field $\Phi$ in modes defined by a multi-index $\vrep=(j_1,...,j_r) \in {\hat{G}}^r$, 
the set of irreducible representation labels of the Cartesian product $G^r$,
and the  ordinary Fourier modes 
of  $\mathbb{R}^d$
labeled by $\vp=(p_1,...,p_\ld)$
as
\begin{equation}\label{eq:Fouriertransform}
    \Phi(\vx,\vg)=\int_{\R^\ld} \frac{\extd\vp}{(2\pi)^{\ld/2}} \text{e}^{i\vp\cdot\vx}\sum_{\vrep \in {\hat{G}}^r}\left(\prod_{c=1}^r d_{j_c}\right)\text{tr}_{\vrep}\left[\Phi_{\vrep}(\vp)\bigotimes_{c=1}^r D^{j_c}(g_c)\right],
\end{equation}
wherein $D^{j}(g)$ are the representation matrices on $d_j$-dimensional representation space, the coefficients of which form a countable complete orthogonal basis of $L^2(G)$ according to the Peter-Weyl theorem. 
Thus, the object $\Phi_{\vrep}(\vp)$, that we also denote at times $\Phi(\vp,\vrep)$,  defines the mode expansion 
of the field $ \Phi(\vx,\vg)$. 
Moreover, the fields transform as a rank-$r$ covariant complex (resp. real) 
tensor, i.e., they transform under unitary (resp. orthogonal) 
transformations $U^c:L^2(G)\to L^2(G)$ in each argument individually, which 
we refer to as the tensorial symmetry. 

A generic tensorial field theory action $S[\Phi,\bar\Phi]$, 
as in usual QFT, 
decomposes into the kinetic and interaction terms. The kinetic term is quadratic and expressed
as $(\Phi,\mathcal{K} \Phi)$ 
where $\mathcal{K}$ is a kernel
that involves the Laplace-Beltrami operator
on $\mathbb{R}^\ld$ and on $ G^{\times \rk}$, potentially decoupled, its powers, and possibly a mass coupling. The interacting part 
$S_{\inter}(\Phi,\bar\Phi)$
is a sum of tensor convolutions
called also contractions. 
If the model is real, a simple prescription would use 
orthogonal invariants
for interactions, 
whereas if the model is complex, then the interactions are unitary invariants. 
At this point, we take full advantage of the Cartesian product of the configuration space
$\mathbb{R}^\ld \times G^{\times \rk}$ to set up the theory space of interactions. 
On one hand, the field
$\Phi(\vx,\vg)$ will be considered
local in the first argument $\vx \in \mathbb{R}^\ld$
and, on the other, nonlocal in the second argument $\vg \in G^{\times \rk}$. In other
words, all following convolutions of the field  $\Phi(\vx,\vg)$ 
are always evaluated at a single point $\vx$ of~$\mathbb{R}^\ld$. 
The measure on this sector
is the Lebesgue measure. 
Meanwhile, the convolution of the same copies of the $\Phi(\vx,\vg)$'s will occur
in radically different way: 
a pair of $\Phi(\vx,\vg)$'s
can only be evaluated 
on a proper subdomain of $G^{\times \rk}$ (at the exclusion of the mass term). 
On this sector, we will use
the Haar measure. 
It is particularly  relevant to  consider the so-called tensor invariants which are a  class of combinatorial non-local interactions which lead to $r$-dimensional (pseudo-)manifolds \cite{GurauBook}.
In such a situation, the convolution of the tensors uses specific combinatorial patterns: each convolution  maps to a (bipartite) colored graph~$\gamma$~\cite{Gurau:2011tj}. 
Given a combinatorial pattern $\gamma$, we denote such a convolution
$\Tr_\gamma[\gf,\bar{\gf}]$, 
where $\Tr$, reminiscent of matrix traces, is regarded
as tensor convolution traces.
Each convolution is performed
via the Haar measure which, thus, becomes implicit in the notation~$\Tr$. 

We will be interested in  a class of interactions given by
\bea
 S_{\inter}(\Phi,\bar\Phi) = \int_{\R^\ld}\extd\vx\sum_{\gamma}\lambda_{\gamma}\Tr_\gamma[\gf,\bar{\gf}]
\eea
where all fields are local
on $\mathbb{R}^{\ld}$ and evaluated at $\vx$, and 
$\Tr_\gamma$ records the convolution of the several copies of $\vg$ according to the pattern dictated by the graph $\gamma$, $\lambda_{\gamma}$
are coupling constants associated with different interactions, each of which coined by a given $\gamma$. The sum over $\gamma$ runs over a finite number of graphs. Explicit examples will follow in Sec.~\ref{sec:BETAcycMelLPA}. 
In the following, 
we use a notation mostly for complex tensor fields. 
The case of real tensors can be inferred from that point without difficulty. 

\subsection{The FRG flow equation: the set up}

For the derivation of the Wetterich-Morris equation \cite{Wetterich:1992yh,Morris:1993qb}, we begin with the generating functional 
\begin{equation}\label{eq:generatingfunctional}
    Z[J,\bar{J}]=\text{e}^{W[J,\bar{J}]}=\int\mathcal{D}\Phi\mathcal{D}\bar{\Phi}\; \text{e}^{-S[\Phi,\bar{\Phi}]+(J,\Phi)+(\Phi,J)},
\end{equation}
wherein $W[J,\bar{J}]$ is the Schwinger functional which generates all connected correlation functions. Using 
\renewcommand{\gf}{\varphi}       
\renewcommand{\gfb}{\bar\varphi}
\[
\gf(\vx,\vg) :=  \langle\Phi(\vx,\vg)\rangle  =  \frac{\delta W[J,\bar{J}]}{\delta \bar{J}(\vx,\vg)}
\quad,\quad 
\gfb(\vx,\vg) :=  \langle\bar{\Phi}(\vx,\vg)\rangle  =  \frac{\delta W[J,\bar{J}]}{\delta J(\vx,\vg)} \, ,
\]
the Legendre transform of $W[J,\bar{J}]$,
\[
\gfi[\gf,\gfb] = \sup_{\bar{J},J}\{(\gf,J) + (J,\gf) - \gfc[\bar{J},J]\}
\]
yields the generating functional of the one-particle irreducible correlation functions.

For the implementation of a renormalization scheme, we introduce scale-dependent versions of these functionals. 
This is achieved by adding a scale-dependent cut-off $k$ to the classical action in \eqref{eq:generatingfunctional}, giving the scale-dependent generating functional
\begin{equation}
    Z_{k}[J,\bar{J}]=\text{e}^{W_{k}[J,\bar{J}]}=
    \int\mathcal{D}\gf\mathcal{D}\bar{\gf}\; 
    \text{e}^{-S[\gf,\bar{\gf}]-(\gf,\mathcal{R}_{k}\gf)+(J,\gf)+(\gf,J)}.
\end{equation}
In this way, one obtains the so-called effective average action by a modified Legendre transform
\begin{equation}
     \Gamma_{k}[\gf,\bar{\gf}]=\sup_{J,\bar{J}}\left[(\gf,J)+(J,\gf)-W_{k}[J,\bar{J}]\right]-(\gf,\mathcal{R}_{k}\gf).
\end{equation}
Consider $\Delta_{\mathbb{R}^\ld}$ 
the ordinary Laplacian on $\mathbb{R}^\ld$
and $\Delta^{(c)}_{G}$ the Laplace-Beltrami on $G$. 
The index~$c$ refers to the tensor color 
index on which this operator will act in $\gf(\vx, g_1,..., g_c,..., g_{\rk})$. 
The effective average action assumes the general form
\begin{align}
  \gfk[\gf,\bar{\gf}] & =(\gf,\mathcal{K}_{k}\gf) +\gfk^{\textsc{ia}}[\gf,\gfb], 
  \label{eq:effectiveaction}\\
  \mathcal{K}_{k} &= - Z_{k} \Delta_{\mathbb{R}^\ld}
  -
  \kappa Z_{k}\sum_{c=1}^\rk (\Delta^{(c)}_{G})^\zeta +\mu_{k}, 
  \label{eq:kineticterm}\\
  \gfk^{\textsc{ia}}[\gf,\gfb] & = \int_{\R^\ld}\extd\vx\sum_{\gamma} \cn{\gamma} \Tr_\gamma[\gf,\bar{\gf}],
  \label{eq:interactions}
\end{align}
in which the dependence on the RG scale $k$ is encoded by the effective mass term $\mu_{k}$ and couplings $\cn{\gamma}$ as well as the wave-function renormalization $Z_{k}$.
We have introduced an extra parameter $0<\zeta\le 1$ allowing to model not only standard short-range propagation ($\zeta= 1$) but also long-range propagation (in particular $\zeta= 1/2$) \cite{Fisher:1972} in the non-local degrees of freedom. 
As a consequence, a dimensionful parameter $\kappa$ is needed in front of $(\Delta^{(c)}_{G})^\zeta$ to balance dimensions which we will treat as a fixed coupling constant which does not flow. 
Stated in another way, in this work, we will be interested only in the flow of a single wave-function renormalization $\wfr$ instead of the second one $Z'_k=\kappa\wfr$.
Letting flow both couplings  will enrich the phase diagram and will be postponed to future investigations. 

The flow equation for the effective average action is then given by  \cite{Wetterich:1992yh,Morris:1993qb}, 
\begin{equation}\label{eq:FRGE}
    k\partial_{k} \Gamma_{k}[\gf,\bar{\gf}]=\frac{1}{2}\STr\left[\left(\Gamma_{k}^{(2)}[\gf,\gfb]+\mathcal{R}_{k}\mathbb{I}_2\right)^{-1}\left(k\partial_{k}\right)\mathcal{R}_{k}\right],
\end{equation}
wherein the \enquote{super-trace} $\STr$ is a trace over all field degrees of freedom, local and non-local.

Notice that the regulator function cuts off momenta $(\vp,\vrep)$
of $\mathbb{R}^\ld\times \hat{G}^r$ 
as given by the transform  \eqref{eq:Fouriertransform}.
Thus, we can switch to the momentum space for making explicit the  cut-off modes. 
The spectrum of the group-space Laplacian is given by the Casimir $C_\rep$ on~$\hat{G}$, i.e. for each representation of $G$ labelled by $\rep$.
Thus, the spectrum of the kinetic part is
\[\label{eq:kinetic spectrum}
(\gf,\kin_k \gf) = \int_{\R^\ld} \extd \vp \sum_{\vrep\in\hat{G}^\rk} \wfr\left( p^2 +
\frac{\kappa}{\cs^\js} 
\sum_{c=1}^{r}(C_{\rep_c})^{\zeta} + \m \right)|\gf(\vp, \vrep)|^2,
\]
where $p^2= ||\vp||^2$ denotes the squared moduli of momenta 
up to unessential $2\pi$ factors.
We further abbreviate the spectrum as well as the dimensionful quantities in the group representation part as 
\[
\cas\vrep:=\sum_{c=1}^{r}(C_{\rep_c})^{\zeta} 
\quad , \quad
\kap:=\frac{\kappa}{\cs^\js} \, .
\]
The regulator function has to satisfy the standard properties
\begin{itemize}
    \item recovery of effective action at $k\to 0$: $\lim_{k\to 0}\mathcal{R}_{k}(\vp,\vrep)\to 0$ for fixed $\vp,\vrep$,
    \item recovery of classical action at large momentum (UV) scale $\Lambda$: $\lim_{k\to \Lambda}\mathcal{R}_{k}(\vp,\vrep)\to \infty$,
    \item regularization at small momenta (IR): $\lim_{p^2,\cas\vrep\to 0}\mathcal{R}_{k}(\vp,\vrep)>0$.
\end{itemize}
For a single momentum scale $k$, the regulator is chosen to be function of 
$ k^\ks-p^2-\kap C_\vrep^{(\zeta)}$. 
Throughout this paper we choose the optimized regulator \cite{Litim:2001up} given by 
\begin{equation}\label{eq:regulator}
     \mathcal{R}_{k}(\vp,\vrep)=Z_{k}
     \left(k^\ks -p^\ps -\kap\cas\vrep \right)
     \theta\left(k^\ks -p^\ps -\kap\cas\vrep \right) \, , 
\end{equation}
where $\theta$ is the Heaviside step function, 
which satisfies the standard regulator properties. 
The derivative of the regulator takes the form
\begin{align}
k\partial_k\reg &= \left[\ks \wfr k^\ks + \partial_k\wfr
\left(k^\ks -p^\ps -\kap\cas\vrep \right)\right] \theta
\left(k^\ks -p^\ps -\kap\cas\vrep \right) \nonumber
\\
&=  \wfr k^\ks \left[\ks - {\eta_k} 
\left(1 -\frac{p^\ps}{k^\ks} -\kap\frac{\cas\vrep}{k^{\ks}} \right) \right] \theta
\left(1 -\frac{p^\ps}{k^\ks} -\kap\frac{\cas\vrep}{k^{\ks}} \right)
\label{eq:regulatorderivative}
\end{align}
where the second line is expressed in terms of the anomalous dimension
\begin{equation}
    \eta_{k}\equiv-\left(k \partial_{k}\right)\log Z_{k}.
\end{equation}
Throughout this paper we keep $\eta_k$ in the equations even though we will eventually set it to zero here, in particular not treat the flow equation for $\wfr$.

In closing this section, we  comment on how this setting could be further extended. 
In the context of the FRG analysis of local scalar field theory, recently the phase diagram for models with two contributions to the kinetic term \textit{and} two individual wave-function renormalizations has been studied~\cite{Buccio:2022egr}. 
It could be interesting to import these ideas to the present context and thus study the impact on the phase structure of our model when introducing one wave-function renormalization for the local and one for the non-local degrees of freedom. 
In fact, we have indications that this might be the only way to derive consistent flow equations for the wave-function renormalization in the first place which we leave for future work, though. 
As another possibility, it could be interesting to introduce two separate RG scales for the different types of degrees of freedom from the beginning since they enter differently in the dynamics, see also~\cite{Marchetti:2022igl,Marchetti:2021wp} for a discussion of this matter. 
Such extensions will be instrumental to understand better the notion of scales for such hybrid theories with local and non-local degrees of freedom. We leave the investigation of these two scenarios to future work.

\section{Flow equations in cylic-melonic local-potential approximation}
\label{sec:BETAcycMelLPA}

The  local-potential approximation (LPA)  already captures important and generic features as those of more general theories. 
This section details the technical 
steps for reaching the beta functions within the LPA. 

\subsection{LPA for non-local interactions}

A tensorial version of the LPA has proven~\cite{Pithis:2020sxm,Pithis:2020kio} to be a valuable tool to understand general features of the tensorial FRG to arbitrary order in the fields.
The LPA is the zero'th order in the derivative expansion of the effective average action $\gfk$.
For a purely local field $\gf(\vx)$ this can be written as the expansion of the integrand 
\[\label{eq:derivativeexpansion}
\gfk[\gf] = \int_{\R^\ld} \extd\vx \left[U_k[\gf(\vx)] +\frac{1}{2}Z_k[\gf(\vx)](\partial\gf)^2(\vx) + \frac{1}{4}Y_k[\gf(\vx)](\partial\gf^2)^2(\vx) + \mathcal{O}(\partial^4) \right] 
 \]
where the zero'th order is described by a potential $U_k$ which is local in a two-fold way: first, by definition it has no derivatives; second, like the full action it contains only point-like interactions (all fields evaluated at the same configuration-space point $\vx$).
For \emph{both} these reasons, the local potential $U_k$ is completely determined by evaluating $\gfk$ at a uniform field configuration $\gf(\vx)=\cgf$,
\[
\gfk[\cgf] = U_k[\cgf] \int_{\R^\ld} \extd\vx = \lv^\ld U_k[\cgf] 
\]
where $\lv$ is a formal expression for the volume of $\R$ (which could be defined by regularization but drops out in the eventual FRG equation anyway \cite{Delamotte}).

Back to our setting, 
adding the group degrees of freedom with tensorial interactions, the zero'th order in the derivative expansion does not contain derivatives. Thus it is local in this sense. However, as it  keeps its previous combinatorial
convolutions, it is not point-like and therefore is  combinatorially non-local.
There is in general no single function $U_k$ to be integrated over. Expanding in field monomials, each term of order~$\gf^{2n}$ contains $\rk\cdot n$ integrations (because each $\rk$-coloured graph $\gamma$ with $2n$ vertices has $\rk n$ edges).
Hence, in the projection to uniform field configurations $\cgf$, we obtain 
\[
\gfk[\cgf] 
= \int_{\R^\ld} \extd\vx \sum_{\gamma}\cn{\gamma}\Tr_\gamma[\cgf]
= \lv^\ld \sum_{n=0}^\infty \Bigg(\sum_{\gamma;|\gamma|=2n} \cn{\gamma} \Bigg) (\cs^\rk \cgf^2)^n
\;, 
\]
where 
the group volume $\cs$ pairs with $\cgf$ into the natural variable 
\[
\rho := \cs^{\rk}\,\cgf^2 \,
\]
We note that the combinatorial non-local information, that is the difference between interactions~$\gamma$ with the same number of vertices $|\gamma|=2n$, is washed out.
Crucially, the zero'th order of the derivative expansion is thus not fully determined by evaluating $\gfk$ at uniform field configurations.

Still, it is meaningful to consider a uniform-field projection also in tensorial theories for two reasons:
first, there are indications that at large $N_k=\cs k$ so called cyclic-melonic interactions dominate \cite{Carrozza:2016tih};
  second, on the right-hand side in the FRG equation there remains still crucial non-local information since it contains the second derivative $\gfk^{(2)}$, see also~\cite{Pithis:2020sxm,Pithis:2020kio} for a detailed discussion.
We will therefore consider here a uniform-field projection on the truncation of $\gfk$ to cyclic-melonic interactions, called the cyclic-melonic potential approximation \cite{Pithis:2020sxm,Pithis:2020kio}.
Even being an approximation to the LPA in the sense of the zero'th order of the derivative expansion, it gives interesting insights into the FRG of tensorial theories at arbitrary order.
In particular, we will show in the following how it helps to understand how local and non-local degrees of freedom interplay in the FRG flow. 

\

\subsection{FRG equation for a cyclic-melonic potential}
\label{subsec:FRGcycMelLPA}

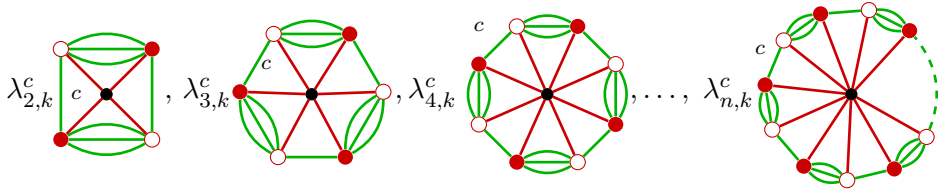
\begin{figure}
\centering
\begin{tikzpicture}
\begin{scope}[xshift=-1cm]
\node [v]       (v)     at (0,0)    {};
\foreach \i in {1,2}{
	\begin{scope}[rotate=\i *180]
	\node [wv]		(w\i)	at (-.6,.6)	{};
	\node [bv]		(b\i)	at (.6,.6)	{};
	\end{scope}
	}
\foreach \i in {1,2}{
	\path	(w\i) edge [eb] 		node 	{}	(b\i)
		(w\i)	edge [eb,bend left=30]node  {}	(b\i)
		(w\i)	edge [eb,bend right=30]node {}	(b\i)
		(w\i)   edge [e]            node    {}  (v)
		(b\i)   edge [e]            node    {}  (v);
	}
\foreach  \i/\j in {1/2,2/1}{
	\path (w\i) edge [eb] node {} (b\j);
	}
\node (c) at (-.4,0) {\scriptsize{$c$}};
\node (l) at (-1,0) {$\cn{2}^c$};
\end{scope}
\begin{scope}[xshift=1.7cm]
\node [v]       (v)     at (0,0)    {};
\foreach \i in {1,2,3}{
	\begin{scope}[rotate=\i *120]
	\node [wv]		(w\i)	at (-.5,.8)	{};
	\node [bv]		(b\i)	at (.5,.8)	{};
	\end{scope}
	}
\foreach \i in {1,2,3}{
	\path	(w\i) edge [eb] 		node 	{}	(b\i)
		(w\i)	edge [eb,bend left=30]node  {}	(b\i)
		(w\i)	edge [eb,bend right=30]node {}	(b\i)
		(w\i)   edge [e]            node    {}  (v)
		(b\i)   edge [e]            node    {}  (v);
	}
\foreach  \i/\j in {1/2,2/3,3/1}{
	\path (w\i) edge [eb] node {} (b\j);
	}
\node (c) at (-.6,.4) {\scriptsize{$c$}};
\node (l) at (-1.5,0) {, $\cn{3}^c$};
\end{scope}
\begin{scope}[xshift=4.8cm]
\node [v]       (v)     at (0,0)    {};
\foreach \i in {1,2,3,4}{
	\begin{scope}[rotate=\i *90]
	\node [wv]		(w\i)	at (-.4,.9)	{};
	\node [bv]		(b\i)	at (.4,.9)	{};
	\end{scope}
	}
\foreach \i in {1,2,3,4}{
	\path	(w\i) edge [eb] 		node 	{}	(b\i)
		(w\i)	edge [eb,bend left=30]node  {}	(b\i)
		(w\i)	edge [eb,bend right=30]node {}	(b\i)
		(w\i)   edge [e]            node    {}  (v)
		(b\i)   edge [e]            node    {}  (v);
	}
\foreach  \i/\j in {1/2,2/3,3/4,4/1}{
	\path (w\i) edge [eb] node {} (b\j);
	}
\node (c) at (-.9,.9) {\scriptsize{$c$}};
\node (l) at (-1.6,0) {,\,$\cn{4}^c$};
\end{scope}
\begin{scope}[xshift=8.8cm]
\node [v]       (v)     at (0,0)    {};
\foreach \i in {1,2,3,4,5}{
	\begin{scope}[rotate=\i *63]
	\node [wv]		(w\i)	at (1.1,.3)	{};
	\node [bv]		(b\i)	at (1.1,-.3)	{};
	\end{scope}
	}
\foreach \i in {1,2,3,4,5}{
	\path	(w\i) edge [eb] 		node 	{}	(b\i)
		(w\i)	edge [eb,bend left=30]node  {}	(b\i)
		(w\i)	edge [eb,bend right=30]node {}	(b\i)
		(w\i)   edge [e]            node    {}  (v)
		(b\i)   edge [e]            node    {}  (v);
	}
\foreach  \i/\j in {1/2,2/3,3/4,4/5}{
	\path (w\i) edge [eb] node {} (b\j);
	}
\path	(w5) edge [eb, dashed, bend right=30]	node {} (b1);
\node (c) at (-1.2,.6) {\scriptsize{$c$}};
\node (s) at (-2.1,0)	{$,\dots, \;\cn{n}^c$};
\end{scope}
\end{tikzpicture}
\caption{Cyclic-melonic interaction vertices diagrammatically described by bipartite $\rk$-colourable vertex graphs (green edges, red vertices, $\rk=4$ in the example) with a distinguished edge colour $c\in\{1,2,...,\rk\}$. }\label{fig:cyclicmelonic}
\end{figure}

Cyclic-melonic interactions are cycles of open melons (see Fig.~\ref{fig:cyclicmelonic}).
For fields on $\R^\ld \times G^\rk$ the calculation of the Hessian $\gfk^{(2)}$ is a straightforward generalization of the purely non-local case $\ld=0$ treated in \cite{Pithis:2020sxm,Pithis:2020kio}.

An open melon of colour~$c$ is a convolution of all group variables of two fields $\gf$ and $\gfb$ except for one, thus given by an operator $\mop$ on $\R^\ld\times G$ with kernel
\[
\mop(\vx,g_c;\vy,h_c) 
:= \delta(\vx-\vy)\int \Bigl(\prod_{b\ne c}\extd g_b\Bigr)\, \gfb(\vx,g_1,...,g_c,...,g_\rk)\gf(\vy,g_1,...,h_c,...,g_\rk ) 
\]
where $\delta(\vx-\vy)$ is the Dirac delta distribution. 
We will also need the operator $\nop$ which convolutes only a single $g_c\in G$, that is the operator $\R^\ld \times G^{\rk-1}$ with kernel
\[
\nop(\vx,\hat\vg_c;\vy,\hat\vh_c) := \delta(\vx-\vy)\int \extd g_c\, \gfb(\vx,g_1,...,g_c,...,g_\rk)
\gf(\vy,h_1,...,g_c,...,h_\rk)
\]
using the notation $\hat\vg_c = (g_1,...,g_{c-1},g_{c+1},...,g_\rk)$. 
Powers of these operators are, as usual, given by convolutions of their kernels, for example
\[
\mop^2(\vx,g_c;\vx',g'_c) = \int_{\R^\ld} \extd \vy \int \extd h_c \mop(\vx,g_c;\vy,h_c)\mop(\vy,h_c;\vx',g'_c) .
\]
One  notes that all convolutions in local variables  of the kind presented just above are straightforward: 
they 
simply deliver one overall delta distribution that involves the external conserved position data.
Furthermore, we understand a zero exponent to yield the identity, that is
\begin{align}
\mop^{0}(\vx,g_c;\vy,h_c) &\equiv \delta(\vx-\vy) \dg{c} \;,  \\
\nop^{0} (\vx,\hat\vg_c;\vy,\hat\vh_c)&\equiv  \delta(\vx-\vy)\prod_{b\ne c}  \dg{b} .    
\end{align}
With this notation, the theory space of cyclic-melonic interactions, as described by the effective average action, is explicitly given by
\[
\Gamma_k^{\textsc{IA}}[\gfb , \gf] 
= \sum_{c=1}^\rk \int_{\R^\ld} \extd\vx \int_{\R^\ld} \extd\vy \int_G \extd g_c 
\; V^c_k\mop(\vx,g_c;\vy,g_c) 
\]
where 
the potential is determined by a single function with expansion as an exponential formal power series
\[\label{eq:potentialfunction}
V^c_k ( z ) = \sum_{n = 2}^{\infty} \frac{1}{n!} \cnk{n} z^n , 
\]
with real scale-dependent coefficients $\cnk{n}$. 
{Physically relevant regimes of the theory are only characterized by a finite number of these coefficients scaling $k^{\theta_n}$ with non-negative powers $\theta_n\ge0$ such that this series is actually convergent.}

To compute the Hessian, we start with the diagonal entries
\[\label{eq:Hessiandiagonal}
\frac{\delta^2\Gamma_k^{\textsc{ia}}[\gf,\gfb]}{\delta\gf(\vx,\vg)\delta\gfb(\vy,\vh)} 
= \delta(\vx-\vy)\sum_{c=1}^\rk \sum_{n=2}^\infty  \frac{n}{n!} \cnk{n} \sum_{l=0}^{n-1}  \mop^{l}(\vx,g_c;\vx,h_c)
\nop^{n-l-1}(\vx,\hat\vg_c;\vx,\hat\vh_c).
\]
Being faithful to the
LPA, we now project to a constant field configuration $\gf(\vx,\vg) = \cgf$, $\gfb(\vx,\vg) = \cgfb$ and define
\[ 
\rho := \cs^\rk \cgfb\cgf
\]
such that on these constant field configurations
\[
\mop^l = \cs^{-1} \rho^l \quad \textrm{and} \quad
\nop^l = \cs^{-(\rk-1)} \rho^l 
\]
which allows to express \eqref{eq:Hessiandiagonal} in terms of derivatives of the potential functions $V_k^c$,
\renewcommand{\dg}[1]{\cs{\delta(g_{#1},h_{#1})}}
\begin{eqnarray}\label{eq:Hessian}
&& \delta(\vx-\vy)\sum_{c=1}^\rk \sum_{n=2}^\infty  \frac{n}{n!} \cnk{n} \cs^{-\rk} \left(\prod_{b\ne c}\dg{b} + \dg{c} + n-2\right)\rho^{n-1} \nonumber\\
&=& \delta(\vx-\vy) \cs^{-\rk} \sum_{c=1}^\rk \left[\left(\prod_{b\ne c}\dg{b} + \dg{c} - 1\right){V_k^c}'(\rho)  + \rho {V_k^c}''(\rho)\right] .
\end{eqnarray}
Consequently, after projection on $\cgf$, all combinatorially non-local information of the cyclic-melonic interactions is contained in the operator
\[\label{eq:deltaoperatorg}
\mathcal{O}^c(\vg,\vh) := \prod_{b\neq c }\dg{b} + \dg{c} - 1 \, .\]
At this point, 
we  emphasize that a LPA in a different regime (e.g. cyclic necklaces \cite{Carrozza:2017vkz}, or any different tensor-invariant
schemes from cyclic-melonic ones), this operator would encode the specific non-locality of that
LPA regime while the rest of the calculations would be essentially the same.

The off-diagonal terms of the Hessian present
no derivatives with respect to neighbouring fields and thus no Dirac delta functions occur. We obtain 
\bea
&&
\frac{\delta^2\Gamma_k^{\textsc{ia}}[\gf,\gfb]}{\delta\gf(\vx,\vg)\delta\gf(\vy,\vh)} 
 = \crcr
 &&
 \delta(\vx-\vy) \sum_{c=1}^\rk \sum_{n=2}^\infty  \frac{n}{n!} \cnk{n} \sum_{p=0}^{n-2}
\int \extd g'\, \mop^{p}(\vx,h_c;\vx,g') \gfb(\vx;g_1,...,g',...,g_\rk)  \crcr
&&  \times\int \extd h'\, \mop^{n-p-2}(\vx,g_c;\vx,h') \gfb(\vx;h_1,...,h',...h_\rk)  
\crcr
&& 
\overset{\gf(\vx,\vg)=\cgf}{\longrightarrow} 
\delta(\vx-\vy) \cs^{-\rk}\,\cgfb^2\,\sum_{c=1}^\rk {V^c}''(\rho) 
\eea
and, respectively for complex-conjugate fields,
\[
\frac{\delta^2\gfk^{\textsc{ia}}[\gf,\gfb]}{\delta\gfb(\vx,\vg)\delta\gfb(\vy,\vh)}
\overset{\gf(\vx,\vg)=\cgf}{\longrightarrow}
\delta(\vx-\vy) \cs^{-\rk}\,\cgf^2\, \sum_{c=1}^\rk {V^c}''(\rho) \,.
\]
To evaluate the full trace $\STr$ in the FRG equation \eqref{eq:FRGE}, this Hessian matrix must be invertible both with respect to the variables $\vx$ and $\vg$ as well as the $2\times2$ structure of the complex field $\gf, \gfb$.
In the momentum space where all operators are diagonal, and according to \eqref{eq:Fouriertransform}, this inversion performs quite simply. 

Under Fourier transformation to momentum space, Dirac delta functions in position space become diagonal in momentum space while constant functions in position space lead delta functions peaking at zero momentum.
The Fourier transform of 
 the non-local operator \eqref{eq:deltaoperatorg} takes the form 
\[\label{eq:deltaoperator}
\mathcal{O}^c_{\vrep} =  \delta_{0\rep_c} +  \prod_{b\neq c }\delta_{0\rep_b} - \delta_{0\vrep} 
\quad \textrm{with} \quad
\delta_{0\vrep}  := 
\prod_{c=1}^\rk \delta_{0\rep_c} \; .
\]
Using all $2\times2$ entries of the Hessian, we evaluate the inverse $(\gfk^{(2)}+\reg)^{-1}$ in the FRG equation \eqref{eq:FRGE} using $\pr:=\wfr(p^\ps+\kap\cas\vrep)+\mu+\reg$ for the Gaussian part which yields
\begin{align}
\left( \pr \id_2 + \gfk^{\textsc{ia}(2)} \right)^{-1}
&=\frac{2\left(\pr+\sum_c \mathcal{O}^c_\vrep {V^c_k}'(\rho)\right)}
{\left(\pr+\sum_c \mathcal{O}^c_\vrep {V^c_k}'(\rho)+\delta_{0\vrep}\rho\sum_c {V^c_k}''(\rho)\right)^2 - \delta_{0\vrep}\left(\rho \sum_c {V^c_k}''(\rho)\right)^2} \nonumber\\
&= \frac{1}{\pr+\sum_c \mathcal{O}_\vrep^c {V^c_k}'(\rho)}
+ \frac{1}{\pr+\sum_c \mathcal{O}_\vrep^c {V^c_k}'(\rho)+ 2\delta_{0\vrep} \rho  \sum_c  {V^c_k}''(\rho)} \, .
\end{align}
Thus, the FRG equation for the cyclic-melonic LPA potential 
\[\label{eq:effectivepotential}
\pot(\rho):= \m \rho + \sum_{c=1}^{\rk} V_k^c (\rho)
\]
becomes
\[\label{eq:Wetterichprojected}
k \partial_k \pot(\rho) =
\frac{1}{2} \int_{\R^\ld}{\extd\vp}
\sum_{\vrep\in{\hat{G}}^{\rk}}\bigg[
\frac{\gp k\partial_k \reg(\vp,\vrep)}{\pr+\sum_c \mathcal{O}_\vrep^c {V^c_k}'(\rho)}
+ \frac{\gp k\partial_k \reg(\vp,\vrep)}{\pr+\sum_c \mathcal{O}_\vrep^c {V^c_k}'(\rho)+ 2 \delta_{0\vrep}  \rho \sum_c  {V^c_k}''(\rho)}
\bigg] \, .
\]
Note that, as detailed in \cite{Pithis:2020sxm,Pithis:2020kio}, for real tensor fields 
instead of complex scalar fields, one arrives at the same equation but only with the second term.

Using the optimized regulator \eqref{eq:regulatorderivative}, the FRG equation is the explicit integro-summation 
\begin{align}\label{eq:WetterichLitim}
\frac{k \partial_k \pot
}{\wfr k^\ks} 
&= \frac{1}{2} \int_{\R^\ld} {\extd\vp }
\sum_{\vrep\in{\hat{G}}^{\rk}} \sum_{\epsilon=0,1}
 \frac{\ks-{\eta_k}\left(1-\frac{p^2}{k^\ks} -\frac{\kap}{k^\ks}\cas\vrep\right)}
{\wfr k^\ks + {\m +   \sum_c \mathcal{O}_\vrep^c {V^c_k}'
+ 2\rho \epsilon \delta_{0\vrep} \sum_c {V^c_k}''
}
}
\theta\left(k^\ks - p^\ps - \kap\cas\vrep \right)
\end{align}
where we introduce $\epsilon=0,1$ to uniformly encode the case of 
real ($\epsilon=1$) and complex fields ($\epsilon=0,1$). 
In this way, it is apparent that the cut-off affects the sum over the spin-labels as well as the the integral over the continuous local momenta.
 Once again, we can comment on the type of LPA regime used: \eqref{eq:WetterichLitim} will keep its form while
the explicit form of the operator $\mathcal{O}_\vrep^c$ might change from one regime
to the other.
In the following, we will consider these combinations of integrals and sums in the specific case of $G=\text{U}(1)$.

\newpage

\subsection{Non-autonomous FRG equation and beta functions}
\label{subsec:SpeccycMelLPA}
\renewcommand{\cas}[1]{\rep^\js}

The compactness of one sector of the configuration space $\R^{\ld}\times 
G^r $ leads to a non-autonomous FRG equation. 
For simplicity, let us consider from now on the ``isotropic'' sector of the theory setting $V^c_k=V_k/\rk$ and thus $\cnk{n} = \cn{n}/\rk$ for all $c=1,...,\rk$: 
this means that we identify all couplings associated with the same melonic interaction up to color symmetry.
This reduces the effective potential \eqref{eq:effectivepotential} 
to 
\[\label{eq:isopotential}
\pot = \m\rho+V_k(\rho) = \m\rho + \sum_{n=2}^\infty \frac{1}{n!} \cn{n} \rho^n  \,.
\]
Setting furthermore $G=\U(1)$, one finds the explicit FRG equation by carrying out the integration and summations in \eqref{eq:WetterichLitim} to be 
\[\label{eq:fullFRGE}
\frac{k \partial_k \pot(\rho)}{\wfr k^\ks} = 
\frac{\kw{0}(k)}{\wfr k^\ks+ \pot'(\rho) + 2\rho\,\pot''(\rho)}
+\frac{\cct \kw{0}(k) + \ccu \rk \kw{1}(k)}{\wfr k^\ks+ \pot'(\rho)}
+\ccu\sum_{s=2}^\rk  \binom{\rk}{s}\frac{\kw{s}(k)}{\wfr k^\ks+ \self{s}(\rho)}
\]
where the last term summing over $s=2,...,\rk$ 
occurs for $\rk>1$ but is not present for $\rk=1$.
The potential $V_k$ enters the effective mass
\[\label{eq:effmass}
\self{s}(\rho) := \m + \frac{\rk-s}{\rk} V_k'(\rho) 
\]
with different factors at each order in $s$ due to the multiplicity of the zero modes 
as proven in App.~A of \cite{Pithis:2020kio}.
{Application of the summation formula (A.7) in \cite{Pithis:2020kio} 
is straightforward: $p^2$~integrals behave with respect to 
$j$-summation as constants, $\rep_c^\js$ summations are simply extended by the $p$ integral.}
The $\eta_k$-dependent integrals $\kw{s}$ with fixed local dimension~$\ld$ contain three parts 
as there are three distinct summations in the numerator in the FRG equation, \eqref{eq:WetterichLitim}, 
to be carried out. 
Let us expand their constituents: 
\[\label{eq:fullspectralsum}
\kw{s}(k) = 
\frac{\ks-\eta_k}{2}\ku{\ld,s}(k)
+ \frac{\eta_k}{2 k^\ks} 
\left( \kp{\ld,s}(k) +\kap \kj{\ld,s}(k) \right) \, ,
\]
where the spectral sums (threshold functions) over functions $f:\R^\ld\times\Z^s \to \R$ are sharply cut off by the regulator,
\[\label{eq:thresholdfunction}
 \kf{\ld,s}_{f}(k) := \int_{\R^{\ld}} \extd\vp \sum_{\vrep \in (\mathbb{Z}\setminus\{0\})^{s} }
\theta\left(k^\ks - p^2 - \kap \cas\vrep \right)
f(\vp,\vrep) \, ,
\]
and we set $\kj{\ld,0}:=0$.
We emphasize that in these spectral sums the zero's of 
$\vrep\in\Z^s$ are taken out.

The spectral sums have slightly artificial properties due to the interplay of the continuous functional setup, the sharp cut-off of the optimized regulator and the discrete spectrum in the $\vrep$ variables.
Their large-$k$ asymptotics correspond to the case of continuous spectra 
and scale as $\ku{\ld,s}(k)\sim k^{\ld+s/ \zeta}$ 
and $\kp{\ld,s}(k) \sim \kj{\ld,s}(k) \sim k^{\ks + \ld+s/\zeta}$ (see App.~\ref{App:Dirichlet}).
But for discrete spectra there are further contributions of lower powers. 
The Euler-Maclaurin formula tells us that they can be expressed in general in a polynomial form up to some rest term 
which in turn can be expanded in general as a Laurent series.
This makes sense as they are monotonically increasing for the relevant cases $f(\vp,\vrep)=1, p^2, \rep^\js$ but not continuous due to the interplay of the $\theta$ function regulator and discrete spectrum.
However, physically this is an artifact of the FRG setup%
\footnote{
When the degrees of freedom are discrete, a discrete RG equation would be more appropriate.
}
and one can as well suppress the rest term and only keep the polynomial part as explained in more detail in Sec.~\ref{sec:dimflow}. 

We obtain flow equations for each coupling by expanding the equation around $\rho=0$ and comparing sides at each order $\rho^n$.
The left-hand side of the equation is a formal power series in the projected average field $\rho$, 
\renewcommand{\m}{\mu}
\renewcommand{\cn}[1]{\lambda_{#1}}
\renewcommand{\mr}{\tilde{\mu}}
\renewcommand{\cnr}[1]{\tilde{\lambda}_{#1}}
\[\label{eq:lhsflowequation}
k \partial_k  \pot(\rho) 
= k\partial_k \m \,\rho + \sum_{n=2}^\infty \frac{1}{n!} k \partial_k\cn{n}\,\rho^n \, ,
\]
where still all couplings are functions of $k$ but we suppress the subscript $k$ from now on for better readability. For the right-hand side we need the Taylor expansion around $\rho=0$ of the fraction%
\footnote{The expansion is simply the case of the Faa di Bruno formula (e.g \cite{Flajolet:2009wm}, III.24) for the function $g\circ f$ with
$g(x)=\frac{1}{x}$ using $g^{(l)}(x) = (-1)^l {l!}/{x^{l+1}}$.
}
\[
\frac1{f(\rho)} = \frac1{f(0)} +
\sum_{n=1}^\infty \frac{\rho^n}{n!} 
\sum_{l=1}^n  (-1)^l \frac{l!}{f(0)^{l+1}} B_{n,l}\left((f'(0),f''(0),...,f^{(n-l+1)}(0)\right) \, ,
\]
which is given in terms of partial (exponential) Bell polynomials
\[
B_{n,l}(x_1,x_2,...,x_{n-l+1}) = \sum_{\substack{\sigma\vdash n \\ |\sigma|=l}} \binom{n}{s_1,...,s_n} \prod_{j=1}^{n-l+1} \left(\frac{x_j}{j!}\right)^{s_j} \, .
\]
Therein, the sum is over partitions $\sigma$ of the expansion order $n$ of length $|\sigma|=l$ with multiplicities $s_j$ for each part $j\in\sigma$, that is
\[
\sum_{j=1}^n s_j = l
\quad , \quad
\sum_{j=1}^n s_j \cdot j = n \, .
\]
As a consequence, the $j\in\sigma$ are bounded by $n-l+1$, thus the sums run effectively only up to $j=n-l+1$, accordingly also the products in the Bell polynomial.

In the full FRG equation \eqref{eq:fullFRGE} there is a sum over two types of fractions with
\[
f_1(\rho) = \wfr k^\ks + \m+\frac{\rk-s}{\rk}{V}_k'(\rho) 
\quad
\textrm{ and }
\quad
f_2(\rho) = \wfr k^\ks + \m+{V}_k'(\rho)+ 2{\rho}{V}_k''(\rho) \, .
\]
The $i$'th derivative of the derivative of the potential at $\rho=0$ is $V^{(i+1)}(0)\equiv f_1^{(i)}(0)=\cn{i+1}$ 
such that
\begin{align}
\frac1{f_1(\rho)} &\equiv \sum_{n=0}^\infty \frac{\rho^n}{n!}\bca{n}\left(\m,\frac{\rk-s}{\rk}\cn{i}\right) 
= \frac{1}{\wfr k^\ks +\m} +
\sum_{n=1}^\infty \frac{\rho^n}{n!} 
\sum_{l=1}^n  \left(\frac{\rk-s}{\rk}\right)^l \bca{n,l}(\m,\cn{i}) \, ,
\label{eq:betaexpansion}
\end{align}
with
\[\label{eq:betacoefficients}
\bca{n,l}(\m,\cn{i}) = \frac{(-1)^l l!}{(\wfr k^\ks+\m)^{l+1}} B_{n,l}\left(\cn{2},\cn{3},...,\cn{n-l+2}\right) \, .
\]
Here we denote the order-$n$ coefficients $\bca{n}$ since they are the beta function part of $\textrm{O}(N)$ vector theories stemming from the $U'$ term (cf.~Sec.~\ref{sec:tensors}). 

In the cyclic-melonic tensorial case, the coefficients \eqref{eq:betacoefficients}  are modified by powers of $\frac{\rk-s}{\rk}$. 
Similarly, the expansion of~$f_2$ yields the beta functions $\bcb{n}$ of scalar field theory which come from the second-derivative $U''$-part. 
There we have $(\rho V'')^{(i)}(0) = i\cn{i+1}$ with the consequence that couplings $\cn{i}$ occur with a factor $1+2(i-1)=2i-1$ such that
\[
\frac1{f_2(\rho)} \equiv \sum_{n=0}^\infty \frac{\rho^n}{n!}\bcb{n}\left(\m,\cn{i}\right)  
= \frac{1}{\wfr k^\ks +\m} 
+ \sum_{n=1}^\infty \frac{\rho^n}{n!} \sum_{l=1}^n \bcb{n,l}\left(\m,\cn{i}\right)
\]
with
\[
\bcb{n,l}\left(\m,\cn{i}\right)  = 
  \frac{(-1)^l l!}{(\wfr k^\ks+\m)^{l+1}} B_{n,l}\left(3\cn{2},5\cn{3},...,(2n-2l+3)\cn{n-l+2}\right) \, .
\]
The Bell polynomials $B_{n,l}$ are sums over multinomials $\prod_i \cn{i}^{t_i}$ for which due to the shift in the $i$-indices $f_1^{(i-1)}(0)=\cn{i}$ we now have
\[\label{eq:betagrading}
\sum_{i=2}^{n+1} t_i = l \quad \textrm{and} \quad \sum_{i=2}^{n+1}  t_i \cdot i= n + l.
\]

Putting everything together, we have the $k$-dependent beta functions $\bfn{n}$ for the cyclic-melonic potential approximation
\begin{align}\label{eq:betaCM}
\bfn{n}(\m,\cn{i};k) = \bcb{n}(\m,\cn{i})\kw{0}(k)
&+ \bca{n}(\m,\cn{i})\left(\cct\kw{0}(k) +  \ccu \rk \kw{1}(k)\right) \\
&+\ccu\sum_{s=2}^\rk  \binom{\rk}{s} \bca{n}\left(\m,\frac{\rk-s}{\rk}\cn{i}\right) \kw{s}(k) 
\, . \nonumber
\end{align}
The FRG equations 
\[\label{eq:FRGE order by order}
\frac{k\partial_k\m}{\wfr k^\ks} = \bfn{1}(\mu_k,\cn{i};k)
\quad , \quad
\frac{k\partial_k \cn{n}}{\wfr k^\ks} = \bfn{n}(\mu_k,\cn{i};k)
\quad \textrm{for } n\ge2
\]
{are non-autonomous due to the dependence of $\bfn{n}$ on the scale $k$ captured by the polynomials $\kw{s}(k)$.
At each order $s$ this non-autonomous part factors from the vector-theory beta functions $\bca{n}$ and $\bcb{n}$ but not overall because of the $\frac{\rk-s}{\rk}$ factors.}

Alternatively, one can factor the non-autonomous part 
\[\label{eq:threshold function expansion}
\fr_l(k) := \kw{0}(k)  + \ccu\rk \kw{1}(k) 
+ \ccu\sum_{s=2}^\rk \binom{\rk}{s}\left(\frac{\rk-s}{\rk}\right)^l \kw{s}(k)
\]
at each order $l$ of the expansion in Bell polynomials
\eqref{eq:betaexpansion},
\bea\label{eq:betaCMcompact}
\boxd{
\bfn{n}(\m,\cn{i};k)= 
\bcb{n}(\m,\cn{i})\kw{0}(k)
+\sum_{l=1}^n \bca{n,l}(\m,\cn{i})
\fr_l(k) 
} \, .
\eea
For example, the flow equation at the first three orders ($n=1,2,3$) are
\begin{align}
\frac{k\partial_k \m}{\wfr k^\ks}  &= 
\frac{-\cn{2}}{(\wfr k^\ks + \m)^2} \left(3\kw{0}  + {\fr_1} \right)(k), 
\label{eq:dimfulcouplingsuptoorder31}\\
\frac{k\partial_k\cn{2}}{\wfr k^\ks} &= 
\frac{-\cn{3}}{(\wfr k^\ks + \m)^2} \left(5\kw{0}  + {\fr_1} \right)(k)
+ \frac{2\cn{2}^2}{(\wfr k^\ks + \m)^3} \left(9\kw{0}  + {\fr_2} \right)(k), 
\label{eq:dimfulcouplingsuptoorder32}\\
\frac{k\partial_k\cn{3}}{\wfr k^\ks} &= 
\frac{-\cn{4}}{(\wfr k^\ks + \m)^2} \left(7\kw{0}  + {\fr_1} \right)(k)
+\frac{6\cn{2}\cn{3}}{(\wfr k^\ks + \m)^3} \left(15\kw{0}  + {\fr_2} \right)(k) \nonumber\\
&\quad +\frac{-6\cn{2}^3}{(\wfr k^\ks + \m)^4} \left(27\kw{0}  + {\fr_3} \right)(k).
\label{eq:dimfulcouplingsuptoorder33}
\end{align}
While in the autonomous limit of the FRG equation one can find dimensionless equations already for the full potential $\pot$, the flow equations for individual couplings are still necessary for explicit calculations that are only possible truncating the space of couplings beyond some $\cn{\nmax}$, that is, setting $\cn{i}=0$ for $i>\nmax$. 
Furthermore, they become crucial to understand the full non-autonomous case which we explore in Sec.~\ref{sec:dimflow}.

\section{Phase structure and fixed points}

Analysis of the phase space given by the FRG equations \eqref{eq:FRGE order by order} requires to specify the spectral sums $\kw{s}$ therein. 
In general, they are combinations of integrals and sums, \eqref{eq:thresholdfunction}, that are difficult 
to handle in full generality but they simplify in their asymptotics.
Thus, we will distinguish in this section three descriptions:
1) We consider the simplified case where the tensorial degrees of freedom are not dynamic at all, that is, the case of $\O(N)^\rk$-invariant local field theory.%
\footnote{Note that this case is different from the setting of~\cite{Eichhorn:2013isa,Eichhorn:2014xaa,Eichhorn:2017xhy,Eichhorn:2018ylk,Eichhorn:2018phj,Eichhorn:2019hsa,Castro:2020dzt,Eichhorn:2020sla} 
where tensor degrees of freedom do not propagate but are nevertheless used as the momentum scale for the renormalization group flow. 
There are no additional local degrees of freedom in that setting.}
2) Then, we consider the regime of large $k^\ks/\kap = (\cs k)^\ks/\kappa$ 
which can be seen either as the large-$k$, the $\cs\to\infty$, or the $\kap\to 0$ regime.
3)  Finally, we discuss the phase space of the full non-autonomous equations when all these parameters are finite.

\subsection{Simplified case: $\O(N)^\rk$-invariant local field theory}
\label{sec:tensors}

Let us start with the simplified case that the non-local degrees of freedom $\vrep$ are 
endowed with no dynamics and so independent of the scale $k$.
This means that there is no propagation of these degrees of freedom which simply corresponds to $\kappa = 0$ here.
Note that this is 
reminiscent of 
 the standard case of local field theories with internal symmetry, as common in gauge field theories. Importantly, such a reduction leads to the context of recently developed 
 \enquote{tensor field theories}~\cite{Witten:2016iux,Gurau:2016lzk,Klebanov:2016xxf,Delporte:2018iyf,Harribey:2022esw, Benedetti:2020iku, Delporte:2020rce, Benedetti:2022twd, Benedetti:2020seh} which have strong connections with 
the famous SYK condensed matter model, see for instance~\cite{Rosenhaus:2018dtp}.

Since the representation labels then do not contribute to the RG scale $k$, we simply cut them off at $|\rep_c|\le N_c$.
Thus the model is an $\times_{c=1}^\rk\O(2N_c+1)$ invariant model of a complex-valued scalar field.
We consider the particular case $2N_c+1=N$ for all $c=1,...,\rk$ which is an $\O(N)^\rk$-invariant model \cite{Benedetti:2014qsa}.
To remove this cut-off, we then consider the large-$N$ limit. 
Spectral sums simplify to the usual local case 
\renewcommand{\ks}{2}
\[
\kf{\ld,s}_{f} 
= \int_{p^\ps<k^\ks} \extd\vp \sum_{\substack{\vrep\in(\Z \setminus\{0\})^{s}\\ \rep_c\le N_c}} f(\vp)
= (N-1)^s \int_{p^\ps<k^\ks} \extd\vp f(\vp) 
\overset{f(\vp)=p^\alpha}{=} (N-1)^s \frac{\ld \vol\ld}{\ld+\alpha} k^{\ld+\alpha} \, ,
\]
such that the full threshold function \eqref{eq:fullspectralsum} becomes
\[
\kw{s}(k) = 
\vol\ld\wfr k^{\ld}\left(1-\frac{\eta_k}{2} + \frac{\ld}{\ld+\ps}\frac{\eta_k}{2} \right) (N-1)^s 
= \vol\ld\wfr k^{\ld}\left(1-\frac{\eta_k}{\ld+2} \right) (N-1)^s  \, .
\]
We denote $\vol\ld=\pi^{\ld/2}/\Gamma(\ld/2 +1)$ the volume of the $\ld$-dimensional unit ball. 
With this the FRG equation \eqref{eq:fullFRGE} is a polynomial in $N-1$,
\begin{eqnarray}
\frac{k \partial_k \pot}{\wfr k^\ks} &=& \vol\ld k^\ld \bigg(1-\frac{\eta_k}{\ld+2} \bigg) 
\crcr
&&\times
 \left[
\frac{1}{\wfr k^\ks+ \pot' + 2\rho\,\pot''}
+\frac{\cct 1 + \ccu \rk (N-1)}{\wfr k^\ks+ \pot'}
+\ccu\sum_{s=2}^\rk  \!\binom{\rk}{s} \frac{(N-1)^s}{\wfr k^\ks+ \self{s}}
\right] .
\end{eqnarray}
One has the usual scaling of a local scalar field theory
\renewcommand{\pot}{u_k}
\renewcommand{\self}[1]{m_k^{(#1)}}
\[\label{eq:rescalingTM}
M_k^{(s)}(\rho) = \wfr k^\ks \self{s}(\rhr) \quad , \quad
U_k(\rho) = c\, k^\ld \pot(\rhr) \quad , \quad
\rho = c\,  k^{\ld-\ks} \rhr / \wfr \, 
\]
with constant $c=\vol\ld$,
or equivalently, for the couplings in the expansion of the potential
\[\label{eq:rescalingTM couplings}
\m = \wfr k^\ks \mr
\quad , \quad
\cn{n} = \wfr^{n} k^{\ks n} (c\,k^\ld)^{1-n} \cnr{n} \quad \textrm{for } n\ge2 \, .
\]
Using $U_k'(\rho)=\frac{\partial \rhr}{\partial \rho}\pot'(\rhr) = \wfr k^\ks \pot'(\rhr)$, this yields the dimensionless FRG equation 
\begin{align}
k \partial_k & \pot(\rhr) + \ld \pot(\rhr) - (\ld-2+\eta_k) \rhr\,\pot'(\rhr) \\
&= \left(1-\frac{\eta_k}{\ld+2} \right) \left[
\frac{1}{1+ \pot'(\rhr) + 2\rhr\,\pot''(\rhr)}
+\frac{\cct 1 + \ccu \rk (N-1)}{1+ \pot'(\rhr)}
+\ccu\sum_{s=2}^\rk  \binom{\rk}{s} \frac{(N-1)^s}{1+ \self{s}(\rhr)}
\right] \, . \nonumber
\end{align}
In particular, for $\rk=1$, the last term does not occur and we recover the FRG equation of the $\O(N)$ model, i.e.~$\O(N)$-invariant scalar field theory,  in the LPA$'$ \cite{Berges:2002ga,Litim:2002jz,Codello:2012ec,Codello:2014yfa} (with additional factor~$\ccu$ because the field is complex here). 

The large-$N$ limit can be taken from the full FRG equation also for $\rk\ge2$.
We understand this  as the limit $N-1=2N_c \to \infty$.
Then, note that at order $(2N_c)^\rk$ the right-hand side is $\rho$-independent since $\self{\rk}=\mr \equiv  \mu/\wfr k^\ks$. 
This only gives a contribution to the flow of the physically irrelevant constant part of the potential.
Neglecting this (possibly infinite) constant part, the leading order is $(N-1)^{\rk-1}$. 
This scaling is the usual for melonic interactions and stems from the $(\rk-1)$-fold edges in the melonic interaction graphs, see Fig.~\ref{fig:cyclicmelonic}.
Upon rescaling the factor $\ccu(N-1)^{\rk-1}$ additional to $\vol\ld$ in \eqref{eq:rescalingTM} and $\pot\to\pot/\rk$, the FRG equation in this limit is
\[\label{eq:FRGE TM}
k \partial_k \pot(\rhr) + \ld \pot(\rhr) - (\ld-2+\eta_k)\rhr\pot'(\rhr)
= \frac{1-\frac{\eta_k}{\ld+2}}{1+ \self{\rk-1}(\rhr)}
= \frac{1-\frac{\eta_k}{\ld+2}}{1+ \frac{\rk-1}{\rk}\mr + \pot'(\rhr)} \, .
\]
This is almost the rescaled FRG equation of $\O(N)$-invariant field theory at large $N$, the only difference being the \textit{extra} $\rk$ multiplicity of the quadratic mass term in the potential.
In principle, the effect of such a different weight of the mass flow as compared to the rest of the potential is already known \cite{Pithis:2020sxm,Pithis:2020kio}, though the relation to the dimension is fixed there as $\ld = \rk-1$.
In the case considered here, $\ld$ and $\rk$ are independent parameters for order-$\rk$ tensors living on $\ld$-dimensional space.

\begin{figure}
\includegraphics[width=.44\linewidth]{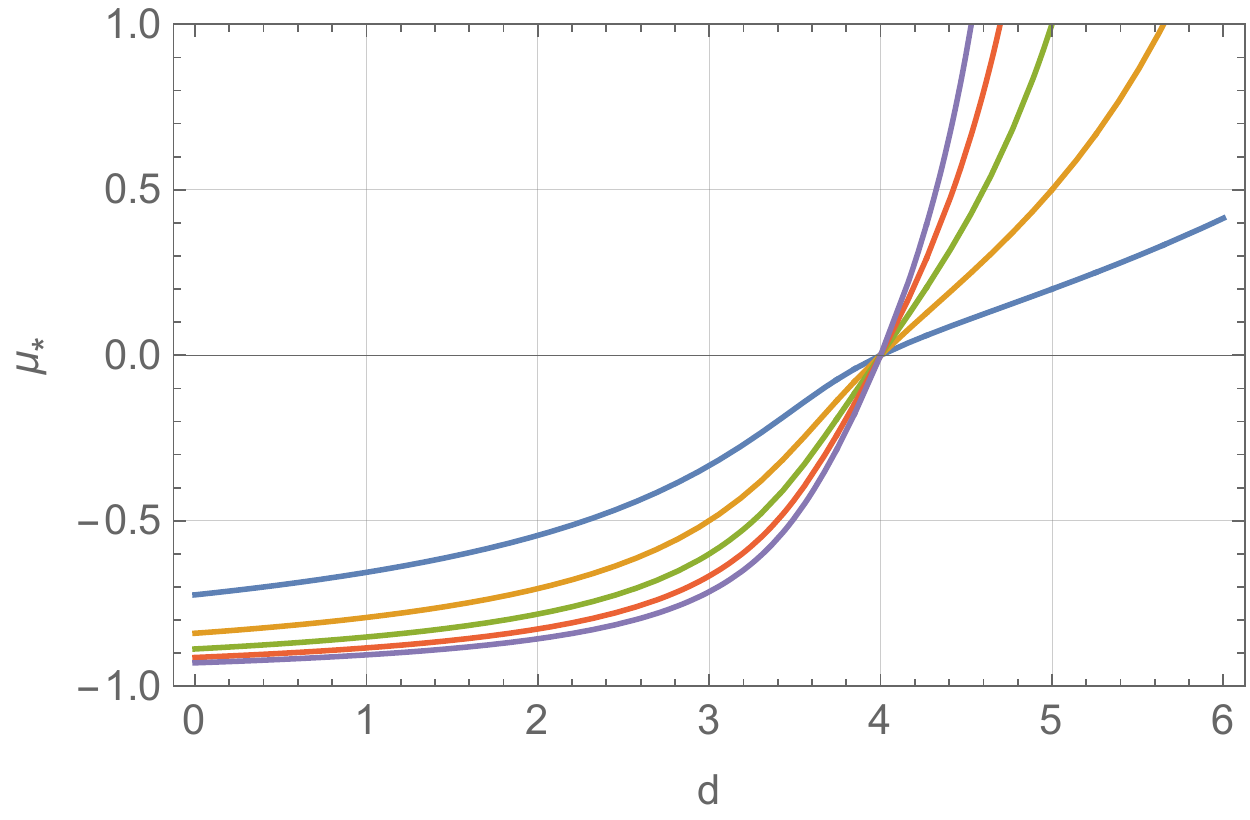}
\includegraphics[width=.52\linewidth]{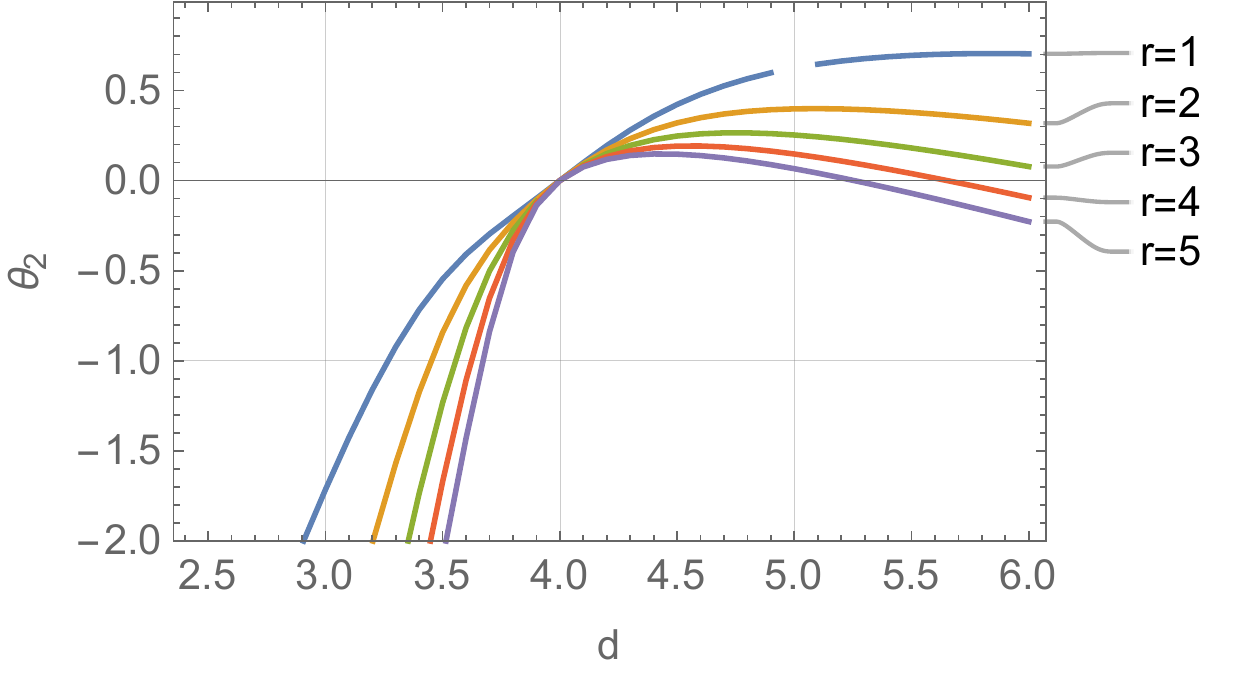}
\centering
\caption{Left: Critical value $\mc=\mc(\ld)$ at the Wilson-Fisher type fixed point at sextic ($n=3$) truncation as a function of the dimension $\ld$ for rank $\rk=1,2,3,4,5$: For any $\rk>0$ this function is bounded from below, $\mc(\ld)>-1$, monotonically increasing up to a neighbourhood of $\mc(4)=0$ which corresponds to the critical dimension being $\crd=4$. With increasing $\rk$ the curve becomes continuously steeper.
Right: Second eigenvalue $\theta_2=\theta_2(d)$ of the stability matrix for the same cases, decreasing with increasing $\rk=1,2,3,4,5$.
}\label{fig:WFP}
\end{figure}

Still, we again find that the phase space of $\O(N)^\rk$-invariant local field theory in the cyclic-melonic LPA ($\eta_k=0$) is qualitatively the same as the one of the vector-model case ($\rk=1$).
In particular, one finds a Wilson-Fisher type fixed point for $2<\ld<\crd = 4$, characterized as the non-Gaussian fixed point continuously connected to the Gaussian fixed point at the critical dimension $\ld=\crd$, with negative $\mc<0$ for $\ld<\crd$ and a single relevant direction, that is, only the largest eigenvalue $\theta_1$ of the stability matrix is positive while all others $\theta_i$, $i\ge2$, are negative for for $\ld<\crd$.
These properties are changed only slightly numerically by the factor $\rk>1$ in the $\O(N)^\rk$-invariant field theory, see Fig.~\ref{fig:WFP} and Tab.~\ref{tab:exponentsd3r3}.

\begin{table}
  \centering
\vspace*{0.5 cm}
\resizebox{0.95\textwidth}{!}{\begin{minipage}{1.25\textwidth}
\centering
\begin{tabular}{|r|c|c|c|c|c|c|c|c|c|c|}
\hline
$\nmax$ & $10\mr$& $10^2\cnr2$& $10^3\cnr3$ & $10^4\cnr4$ & $10^5\cnr5$ & $10^6\cnr6$ & $10^7\cnr7$ & $10^8\cnr8$ & $10^9\cnr9$ & $10^{10}\cnr{10}$ \\ \hline\hline
 6 & -6.5649 & 5.1643 & 9.4342 & 15.067 & 7.9684 & -54.935 & \text{} & \text{} & \text{} & \text{} \\
 7 & -6.5541 & 5.1883 & 9.4629 & 14.916 & 6.0346 & -73.574 & -229.55 & \text{} & \text{} & \text{} \\
 8 & -6.5563 & 5.1834 & 9.4570 & 14.947 & 6.4366 & -69.694 & -181.66 & 797.55 & \text{} & \text{} \\
 9 & -6.5576 & 5.1806 & 9.4538 & 14.964 & 6.6554 & -67.584 & -155.63 & 1230.5 & 8760.4 & \text{} \\
 10& -6.5575 & 5.1808 & 9.4540 & 14.963 & 6.6390 & -67.743 & -157.59 & 1198.0 & 8102.3 & -15350. \\
 11& -6.5573 & 5.1811 & 9.4544 & 14.961 & 6.6164 & -67.961 & -160.28 & 1153.3 & 7198.1 & -36441. \\
 12& -6.5573 & 5.1811 & 9.4544 & 14.961 & 6.6157 & -67.967 & -160.35 & 1152.0 & 7172.4 & -37040. \\
\hline
\end{tabular}
\newline
\vspace*{0.5 cm}
\newline
\begin{tabular}{|r|c|c|c|c|c|c|c|c|c|c|c|c|}
\hline
$\nmax$ & $\theta_{1}$& $\theta_2$& $\theta_3$ & $\theta_4$ & $\theta_5$ & $\theta_6$ & $\theta_7$ & $\theta_8$ & $\theta_9$ & $\theta_{10}$ 
\\ \hline\hline
 6 & 0.50915 & -1.7691 & -5.5429 & -9.9919 & -16.288 & -28.526 & \text{} & \text{} & \text{} & \text{} \\
 7 & 0.51807 & -1.7196 & -4.4455 & -8.5409 & -12.944 & -21.296 & -34.652 & \text{} & \text{} & \text{} \\
 8 & 0.51817 & -1.7601 & -3.9621 & -7.3798 & -11.061 & -17.086 & -26.710 & -41.022 & \text{} & \text{} \\
 9 & 0.51716 & -1.7723 & -3.8661 & -6.5101 & -9.8464 & -14.329 & -21.803 & -32.301 & -47.464 & \text{} \\
 10&0.51704 & -1.7673 & -3.9116 & -6.0278 & -8.9458 & -12.485 & -18.399 & -26.781 & -38.014 & -53.954 \\
 11&0.51714 & -1.7650 & -3.9374 & -5.9025 & -8.2795 & -11.246 & -15.945 & -22.858 & -31.940 & -43.840 \\
 12&0.51716 & -1.7654 & -3.9317 & -5.9493 & -7.8900 & -10.401 & -14.165 & -19.931 & -27.550 & -37.247 \\
\hline
\end{tabular}
\end{minipage}}
 \caption{\label{tab:exponentsd3r3}%
Values of the coupling constants and scaling exponents (eigenvalues of the stability matrix) at the Wilson-Fisher type fixed point for the $\ld=3$ dimensional $\O(N)^{\rk=3}$-invariant local field theory in $(\gfb\gf)^{\nmax}$ truncation.
Convergence with higher orders $\nmax$ justifies to draw conclusions from results at finite $n$.
}
\end{table}

\subsection{Phase structure in the autonomous limit}
\label{sec:UV/large volume}

The above tensor toy model is very similar to the full theory in the regime of large 
\[
\ak := \left( \frac{k^\ks}{\kap} \right)^{\frac{1}{\js}}
= \cs\left(\frac{k}{\sqrt{\kappa}}\right)^{\frac{1}{\zeta}}
\]
but differs in dimension.
When the non-local degrees of freedom $\vrep$ are dynamic ($\kappa>0$), the full FRG equation (\ref{eq:fullFRGE}) is non-autonomous and thus very difficult to handle.
A common strategy is to consider autonomous equations in an appropriate asymptotic limit.
In the theory with local and tensorial degrees of freedom, this is the limit of large $\ak$.

The large-$\ak$ limit has several interpretations.
First, it describes the large-$k$ ``ultraviolet'' (UV) regime of the theory. Thus, it is sufficient for the search of renormalizable fixed points of the theory.
Furthermore, it is also the limit of small $\kap=\kappa/\cs^\js$ which can have two meanings: either the ``Tensor-Model \emph{limit}'' $\kappa\to 0$ or the large-volume, or ``thermodynamic'' limit $\cs\to\infty$~\cite{BenGeloun:2016kw} . 
Certainly, it can also be regarded as the double- or triple-scaling limit of a combination of these.

In the large-$\ak$ limit the spectral sums \eqref{eq:thresholdfunction} can be approximated by integrals over $q^\js = (\rep/\cs)^\js \in [0,k^\ks]$.
These are well known integrals since Dirichlet's times \cite{Dirichlet:1839} as we remind in App.~\ref{App:Dirichlet}.
Then, the sum of all threshold functions \eqref{eq:fullspectralsum} is
\[
\kw{s}(k) \underset{\ak\to\infty}{\sim} 
\frac1 2 \vks{\ld,\rk} \wfr k^{\ld} \ak^s \left(\ks-\eta_k \left(1-\frac{\ld+\frac{s}{\zeta}}{\ld+\frac{s}{\zeta} + \ks}\right) \right)\, , 
\]
where the overall constant is, according to \eqref{eq:generalized volume}, the volume 
\[
\vks{\ld,\rk} = \frac{\pi^{\frac{\ld}{2}} 2^{s}\Gamma\left(1+\frac{1}{\js}\right)^s}{\Gamma\left(1+\frac{\ld}{\ps}+\frac{s}{\js}\right)} \, .
\]
As a consequence, the full FRG equation (\ref{eq:fullFRGE}) becomes polynomial in $\ak$. 
However, as we consider the large-$\ak$ regime, only the order $\ak^{\rk-1}$ is relevant for $\rk>1$ (since the $\ak^\rk$ term contributes again only to the irrelevant constant part of the potential)
and we obtain
\[
\frac{k \partial_k U_k(\rho)}{\wfr k^\ks} = 
\ks \rk \vks{\ld,\rk} \wfr k^{\ld} \ak^{\rk-1} \frac{1-\frac{\eta_k}{\ld+\ks+\frac{\rk-1}{\zeta}}}{\wfr k^\ks+ M_k^{(s)}(\rho)} \, .
\]
Here, the case of $\rk=1$ is special as the dominant threshold function $\kw1$ scales then $~k^\ld \ak$ and there is a factor $\ak$ in the FRG equation, too.

To derive dimensionless flow equations, we need the same rescaling as for the $\O(N)^\rk$-invariant local field theory, \eqref{eq:rescalingTM}, with a constant $c=\ks \vks{\ld,\rk} \kap^{-\frac{\rk-1}{\js}}$,
\[\label{eq:rescalingTFT}
M_k^{(s)}(\rho) = \wfr k^\ks \self{s}(\rhr) \quad , \quad
U_k(\rho) = c\,\rk\, k^\efd \pot(\rhr) \quad , \quad
\rho = c\, k^{\efd-\ks} \rhr /\wfr \, ,
\]
or equivalently, for the couplings in the expansion of the potential
\[\label{eq:rescalingTFT couplings}
\m = \wfr k^\ks \mr
\quad , \quad
\cn{n} = \rk \wfr^{n} k^{\ks n} (c\, k^\efd)^{1-n} \cnr{n} \quad \textrm{for } n\ge2 \, ,
\]
only now with an effective dimension
\[\boxd{
\efd = \ld + \frac{\rk - 1}{\zeta} \, 
}\]
{for $\rk>1$, while for $\rk=1$ this dimension is $\efd=\ld+1/\zeta$.}
Accordingly, the flow equation for $\rk>0$ is
\[\label{eq:FRGE TFT}
k \partial_k \pot(\rhr) + \efd \pot(\rhr) - (\efd-\ks+\eta_k)\rhr\,\pot'(\rhr)
= \frac{1-\frac{\eta_k}{\efd+\ks}}{1+ \frac{\rk-1}{\rk}\mr + \pot'(\rhr)} \, 
\]
which is the same equation as for the $\O(N)^\rk$-invariant local field theory, \eqref{eq:FRGE TM}, but with the modified dimension $\efd$.
This modification of the effective dimension has already been found in~\cite{Marchetti:2021wp} in the Gaussian approximation; thus, our result here is a generalization to the non-perturbative case of the full phase space. 

Since the flow equations of the large-$\ak$ tensorial field theory are the same as for the $\O(N)^\rk$-invariant local field theory, so are the solutions and the resulting phase space.
The only difference is that in the local field theory, the relevant parameters $\ld$ and $\rk$ are completely independent.
Here, in the large-$\ak$ tensorial field theory, we have still independent $\ld$ and $\rk$ but the relevant two parameters in the equation are $\rk$ and the effective dimension $\efd = \ld + (\rk -1)/\zeta$. 
Thus, there is only a limited number of choices for theories below the critical dimension, $\efd < \crd = 4$.
For example, $\efd=3$ is realized for $\zeta=1$ only by $(\ld,\rk)= (2,2),(1,3)$ and $(0,4)$.
These examples are already fully discussed in the previous section on the $\O(N)^\rk$-invariant local field theory: 
There is the Wilson-Fisher type fixed point, only quantitatively modified when $\rk>1$. Here, the magnitude of these modifications is even bounded as $\rk-1<\zeta\efd$ is a necessary condition for this fixed point.
The case of smaller $\zeta<1$, in particular $\zeta=1/2$, has already been discussed in \cite{Pithis:2020sxm,Pithis:2020kio} for $\ld=0$ with the same result of only minor quantitative modifications. 
For additional local dimension $\ld>0$ the parameter $\rk$ becomes only restricted to smaller values and thus there are only smaller numerical modifications of the Wilson-Fisher fixed point also in these cases.

\renewcommand{\pot}{U_k}
\renewcommand{\self}[1]{M_k^{(#1)}}
\subsection{Renormalization group flow in the non-autonomous case}\label{sec:dimflow}

If $\kap = \kappa/\cs^\js$ has a fixed finite value the FRG equation (\ref{eq:fullFRGE}) is truly non-autonomous.
This is the case for compact groups $G$ which have a finite volume $\cs$, in particular here if really $G=\U(1)$ and not just used as a compactification of $G=\R$.
Strictly speaking, it is not possible to obtain then a dimensionless flow equation via rescaling.
However, one can still find a rescaling which shifts the dependence on the scale $k$ completely into the effective dimension $\efd$ \cite{Pithis:2020sxm,Pithis:2020kio}.
Using this trick, we analyse the case of finite $\kap$ in the first part of this Section. In the second we provide numerical solutions to the non-autonomous flow equation corroborating the results.

\

In the FRG equation (\ref{eq:fullFRGE}) there are not only different spectral sums $\kw{s}$ occurring at each order $s=0,1,...,\rk-1$ but there is also a modification of the potential term~$\pot$ by the factor $\frac{\rk-s}{\rk}$ leading to the effective mass terms $\self{s}$.
Only because of this modification it is not possible to factor the $k$-dependent sum over spectral sums $\kw{s}$ from the combinatorial part encoded in the potential.
On the level of beta functions \eqref{eq:betaCMcompact} one can shift this entanglement of combinatorial and $k$-dependent part to the expansion of the Bell polynomials which at the same time is the expansion in the factor $\frac{\rk-s}{\rk}$ given by functions~$\fr_l(k)$, \eqref{eq:threshold function expansion}.
We use the first order $\fr_1$ of this expansion to define a scaling of the couplings; higher orders $\fr_l$ in the expansion then describe simply the flow of the factor $\frac{\rk-s}{\rk}$ from $
\frac1{\rk}$ for $s=\rk-1$ in the UV 
to $\frac{\rk-0}{\rk} = 1$ for $s=0$ in the IR.

\begin{figure}
\includegraphics[width=.48\linewidth]{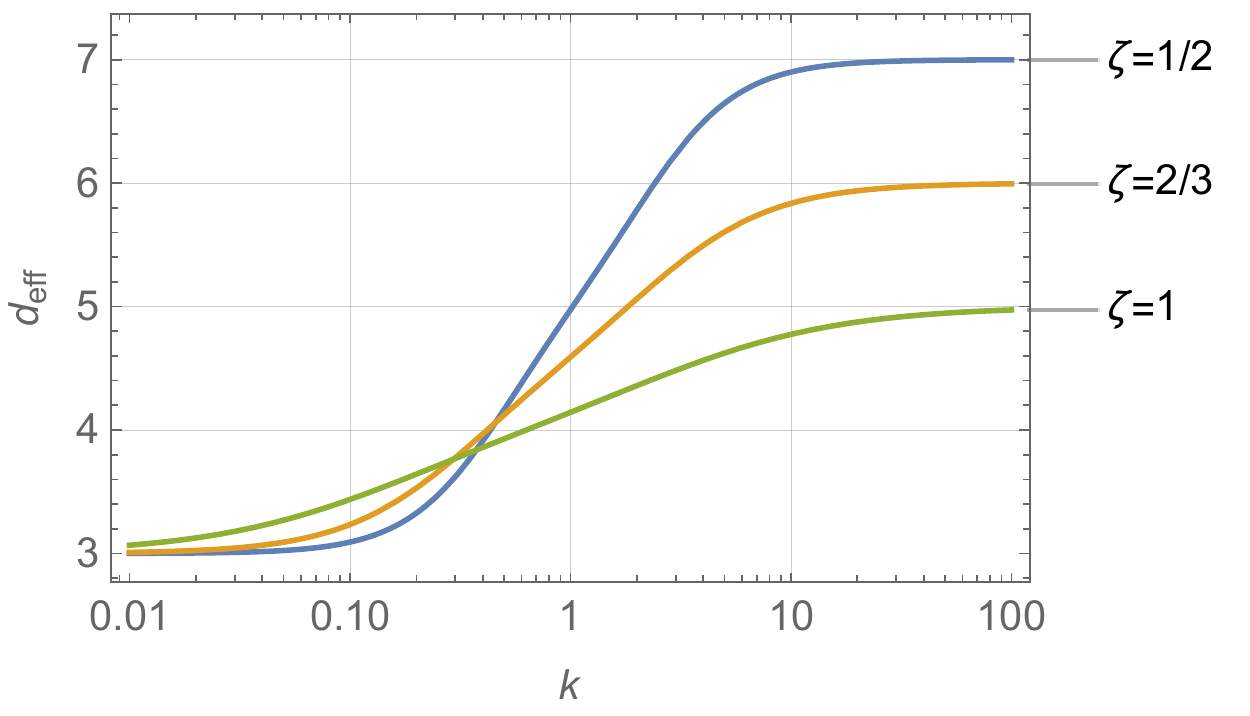}
\includegraphics[width=.4\linewidth]{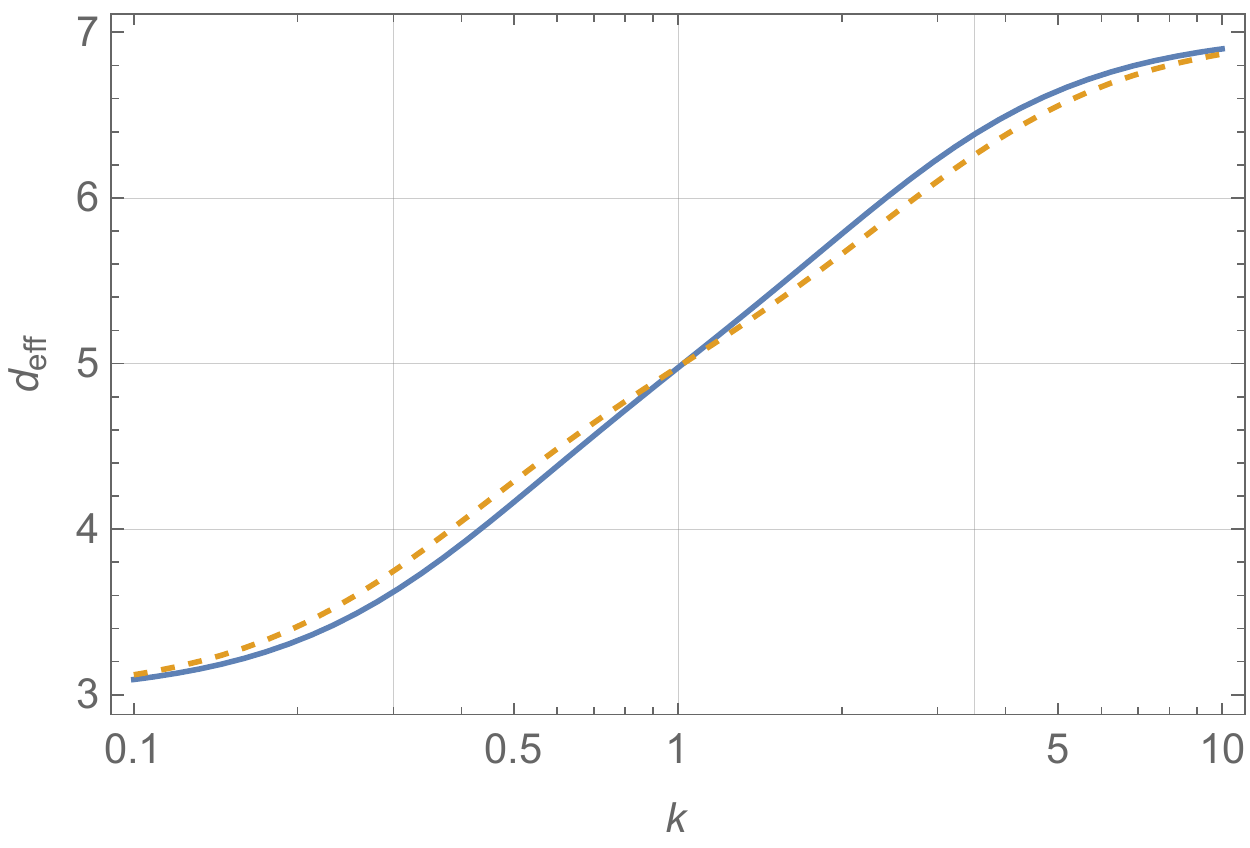}
\centering
\caption{Left: Comparing the flow of effective dimension for different values of $\zeta$ in the case $\ld=\rk=3$ (with $\kap=1$, $\eta_k=0$) 
using the asymptotic approximation \eqref{eq:threshold function asymptotics 1} for the threshold function $\ku{3,3}$.
Right: Comparing for $\zeta=\frac{1}{2}$ the asymptotic approximation (solid curve) with
the full polynomial expression \eqref{eq:threshold function simplex} (dashed); this shows that details of approximation give the same qualitative results, with only minor differences in numerics.
}\label{fig:dimensionflow}
\end{figure} 

Concretely, this rescaling takes the form
\[
\m = \wfr k^\ks \mr
\quad , \quad
\cn{n} = \wfr^{n} k^{\ks n} \fr_1(k)^{1-n} \cnr{n} \quad \textrm{for } n\ge2 \, .
\]
It is the same rescaling as before, \eqref{eq:rescalingTFT couplings}, only changing the power function $c k^{\ld+(\rk-1)/\zeta}$ for the function $\fr_1$.
In fact, for $\eta_k = 0$ the former is exactly the asymptotics of $\fr_1$.
On a finite scale $k$, however, the function $\fr_1$ scales with a power
\[\label{eq:effective dimension}
\efd(k) := k\partial_k \log \fr_1(k)  \, .
\]
Indeed, the full rescaled flow equations become
\[\label{eq:fullnonautonomous}
\boxd{
k\partial_k \cnr{n} 
+\efd(k)\cnr{n} - n(\efd(k) - \ks  + \eta_k )\cnr{n} 
=R_0(k)  \bcb{n}(\mr,\cnr{i})
+\sum_{l=1}^n R_l(k)  \bca{n,l}(\mr,\cnr{i})
}
\]
which shows that the effective scaling $\efd(k)$ is again an effective dimension in the sense that the flow equations are of the type of $\efd(k)$-dimensional $\O(N)$ models,
up to factors 
\[
R_0(k):= \frac{\kw{0}(k)}{\fr_1(k)} 
\quad\text{ and }\quad
R_l(k):= \frac{\fr_l(k)}{\fr_1(k)} \, .
\]
All of the non-autonomous properties of the equation is now encoded in these $k$-dependent factors but these are simple interpolations between the asymptotic regimes:
\begin{itemize}
    \item The first one, $R_0$, captures the fact that the term with second-derivative potential~$\pot''$ occurs only at lowest order in $\kw{0}$ in the FRG Equation (\ref{eq:fullFRGE}). 
    Accordingly, the fraction cancels to $R_0(k\to 0)=1$ at small scales while it goes to $R_0(k\to\infty)=0$ at large scales.
    \item The other ones, $R_l$, capture the relative factor $\frac{\rk-s}{\rk}$ at each order $\kw{s}$ which results in a monotonic flow from $R_k(k\to\infty) = 1/\rk^{l-1}$ at large scales to $R_k(k\to 0)=1$ at small scales. 
    \item Last, and most importantly, the effective dimension $\efd(k)$ flows from $\efd(k\to\infty)=\ld+(\rk-1)/\zeta$ in the UV to $\efd(k\to 0) = \ld$ in the IR.     
\end{itemize}

\begin{figure}
\includegraphics[width=.5\linewidth]{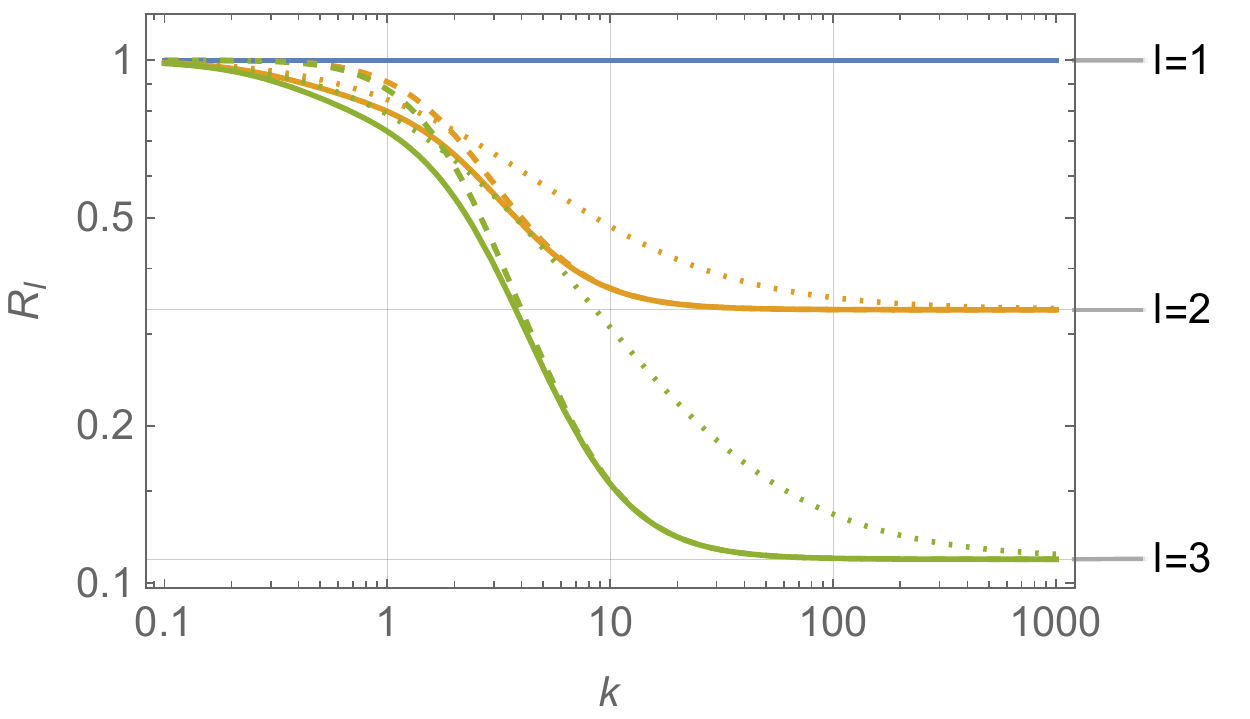}
\centering
\caption{The flow of the factors $R_l$ in the case $\ld=\rk=3$ (with $\kap=1$, $\eta_k=0$) for $l=1,2,3$, comparing for $\zeta=\frac{1}{2}$ the full polynomial expression \eqref{eq:threshold function simplex} (solid curves) with the asymptotic approximation \eqref{eq:threshold function asymptotics 1} (dashed), and for the latter also with the case $\zeta=1$ (dotted). In any case, $R_l$ flow from one in the IR to $1/r^{l-1}$ in the UV.
}\label{fig:Rlflow}
\end{figure} 

The precise behaviour of the functions $\efd$, $R_0$ and $R_l$ depend on the spectral sums~$\kf{\ld,s}$, \eqref{eq:thresholdfunction}, and thereby on the regulator and approximations chosen.
Still, the qualitative behaviour is rather universal and independent of these specific choices, as Fig.~\ref{fig:dimensionflow} and \ref{fig:Rlflow} show.

A subtle case is the small-$k$ limit of the effective dimension $\efd$ since this comes from the contribution of $\kw{s}$ of lowest order in $k$.
As mentioned already when introducing the spectral sums in \eqref{eq:thresholdfunction}, these are in general Laurent series due to the interplay of integration of continuous local momenta and summation of discrete non-local momenta together with the sharp regulator \eqref{eq:regulator}. 
However, the occurrence of monomials in $1/k$ in the expansion should be understood as an unphysical artifact of applying the continuous functional RG method to a theory with discrete momenta. 
In fact, the physically relevant form of the spectral sums is a product $k^{\ld}$ times a polynomial of order $\rk-1$ in $\ak$ which gives the result $\efd(k\to 0) = \ld$.
We discuss the details of the explicit expressions of the spectral sums in App.~\ref{app:threshold}.

In this way, the full and rather non-tractable non-autonomous flow equations allow an interpretation as ``quasi-autonomous'' equations.
They are of a form similar to standard autonomous equations, only with the dimension $\efd$ monotonically interpolating between the asymptotic values $\ld+(\rk-1)/\zeta$ and $\ld$. 
Furthermore, the special $1/\rk$-factor distinguishing the theory at large $\ak$ from $\O(N)$ models at large $N$, as discussed in Sec.~\ref{sec:UV/large volume}, is monotonically removed towards the small-$k$ regime, encoded in the factors $R_l(k)$. 
And, as a last subtlety, in the small-$k$ regime the $\pot''$-term is restored via $R_0$ such that the asymptotic equations in that regime become exactly those of $\U(1)\simeq \O(2)$-invariant field theory in $\ld$ dimensions. Consequently, one yields the autonomous flow equation for the rescaled potential $\pot(\rhr)$,
\renewcommand{\pot}{u_k}
\begin{equation}\label{eq:FRGequationIR}
    k\partial_k \pot(\rhr)+\ld \pot(\rhr)-(\ld-\ks)\rhr\,\pot'(\rhr)
    =\frac{1}{1+\pot'(\rhr)}+\frac{1}{1+\pot'(\rho)+2\rhr\, \pot''(\rhr)} \, ,
\end{equation}
which is that of a local complex-valued scalar
field theory on $\ld$-dimensional Euclidean space~\cite{Berges:2002ga,Delamotte,Dupuis:2020fhh}.
Thus, we record that the theory becomes effectively local in this limit since the tensorial degrees of freedom of the model effectively vanish.

Assuming that the full equations are sufficiently well-behaved such that continuous global solutions for $k\in \R_+$ exists, 
we can  identify fixed points of the full equations in either asymptotic regime.
UV attractive fixed points are identified using the autonomous equations in the limit $k\to \infty$.
This is covered by the limit $\ak\to\infty$ and thus the discussion of the last section applies.
On the other hand, IR attractive fixed points can be determined by the autonomous equations in the limit $k\to 0$ in which we have identified the equation of $\ld$-dimensional scalar field theory, \eqref{eq:FRGequationIR}.
This asymptotic regime is relevant for the question of phase transitions.

According to this interpretation, we expect the system to exhibit a transition between a broken and symmetric phase of the global $\text{U}(1)$-symmetry in the tensorial field theory on $\mathbb{R}^d\times \text{U}(1)^r$ at any rank as long as $d>2$ in accordance with the Mermin-Wagner-Hohenberg theorem~\cite{Hohenberg:1967zz,Mermin:1966fe,Coleman:1973ci} which disallows the spontaneous breaking of continuous symmetries in two or less dimensions. This expectation is also supported by prior studies employing Landau-Ginzburg mean-field theory~\cite{Pithis:2018eaq,Marchetti:2021wp,Marchetti:2022igl} and the FRG methodology~\cite{Benedetti:2014qsa,Benedetti:2015yaa,Lahoche:2016xiq,BenGeloun:2015ej,BenGeloun:2016kw,Pithis:2020sxm,Pithis:2020kio} to TGFT. 

We investigate this issue by numerically integrating the full non-autonomous beta functions~(\ref{eq:fullnonautonomous}) for the dimensionful potential $U_k(\rho)=\mu_k\rho+V_k(\rho)$, 
as defined in~\eqref{eq:isopotential}, from $k =\Lambda$ in the UV to small $k$ in the IR.%
\footnote{To solve the dimensionful flow equations, which form a set of coupled first-order non-linear differential equations, we utilize the Runge-Kutta method at machine precision implemented in Mathematica's NDSolve function.} 
To this aim, we start in the UV with any potential which displays spontaneous breaking of the global $\text{U}(1)$-symmetry. 
Upon integration towards $k=0$, depending on the initial conditions $\mu_{\Lambda},\lambda_{n,\Lambda}$ with $n\geq 2$, one observes two distinct behaviours. 
Either the potential flattens completely out already at a value $k>0$ and thus exhibits a global minimum at $\rho=0$ corresponding to symmetry restoration towards the IR. Alternatively, the potential evens mildly out but still exhibits a non-trivial global minimum corresponding to the effective potential of a broken phase. 

\begin{figure}
\centering
  \includegraphics[width=.311\linewidth]{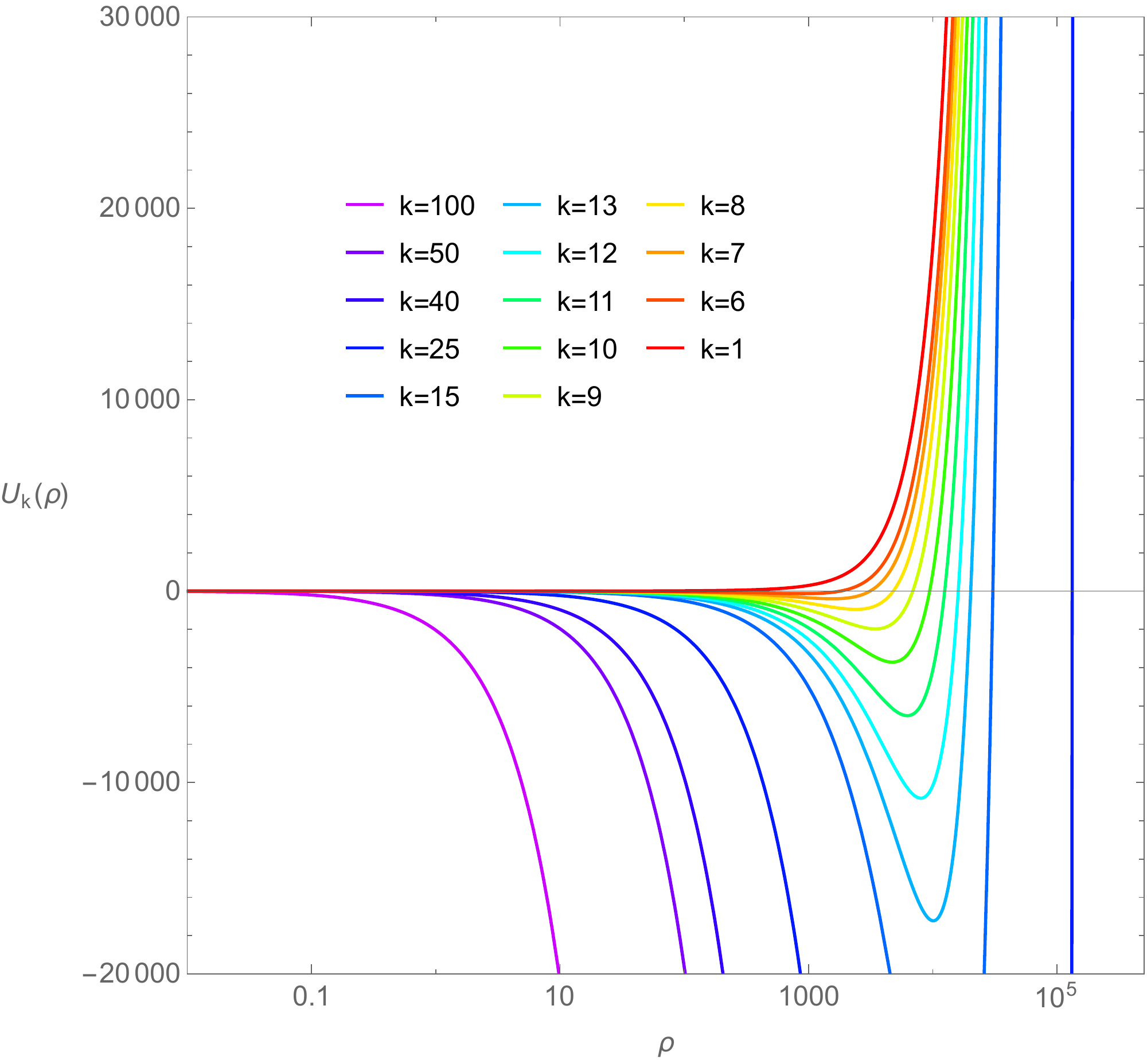}
  \includegraphics[width=.311\linewidth]{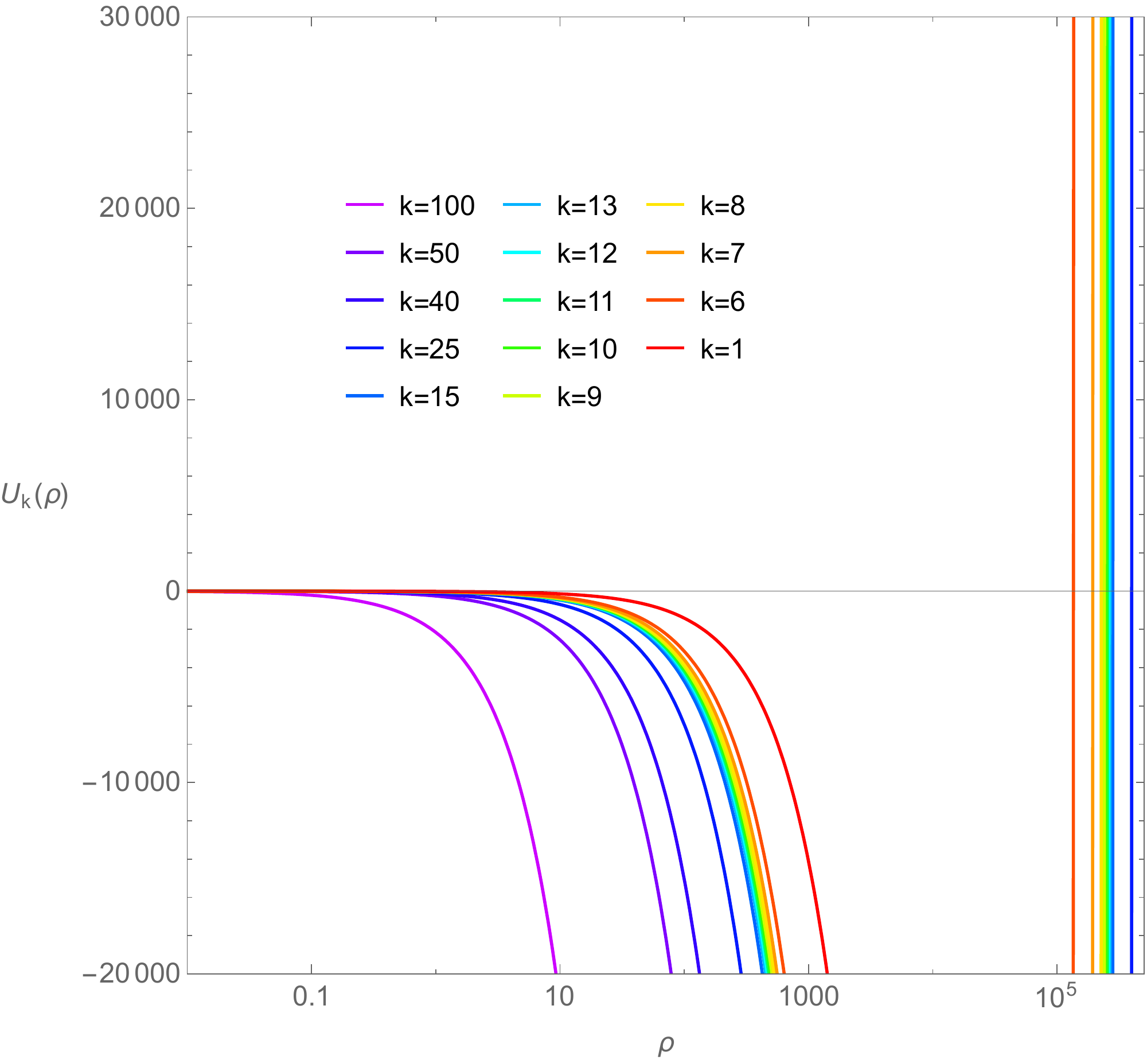}
  \includegraphics[width=.299\linewidth]{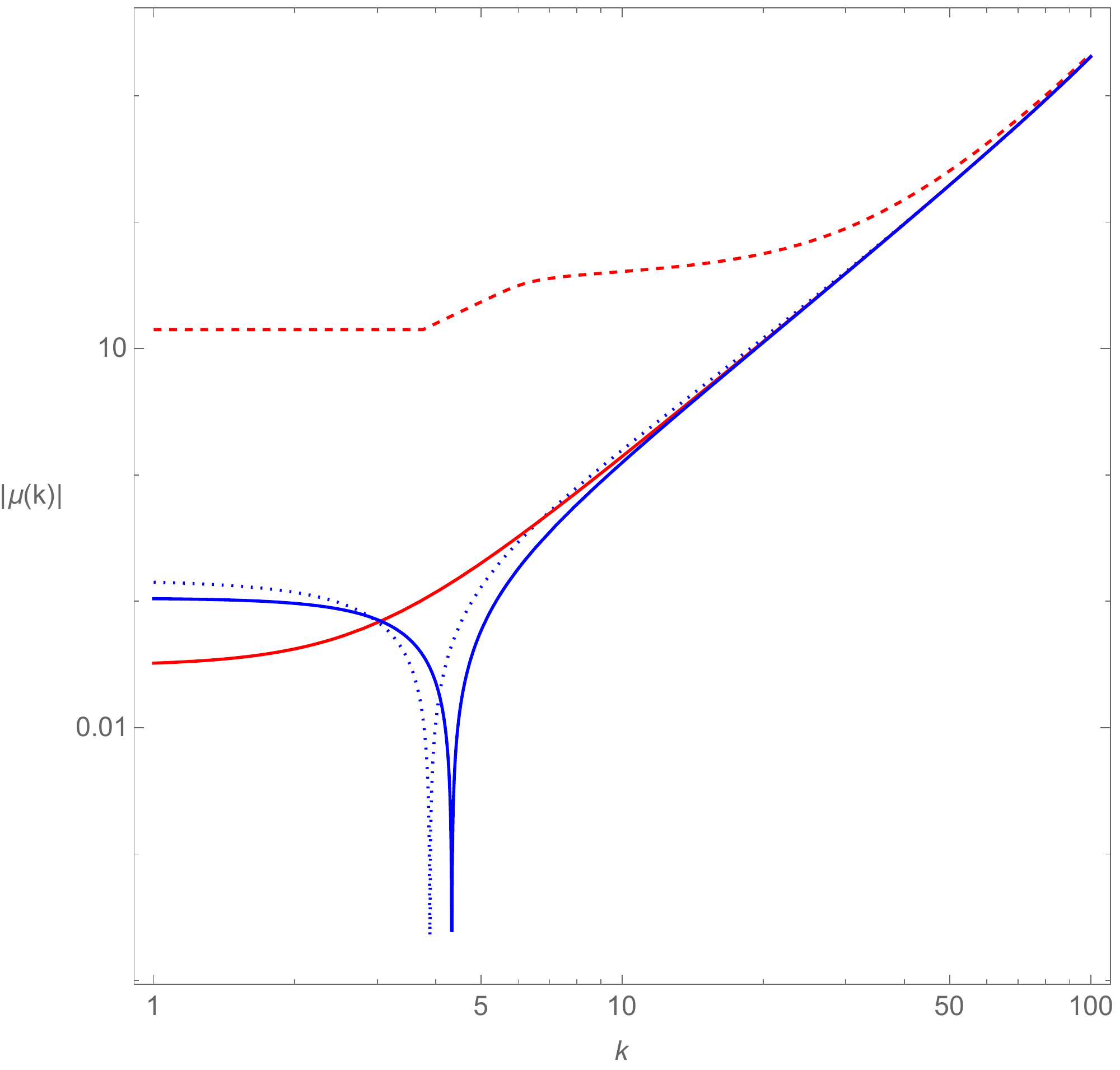}
\caption{Left and center panel: Illustration of the flow of the dimensionful potential $U_k(\rho)$ at rank $\rk=3$ with $d=3$ and $\zeta=1$ in the $\nmax=3$ truncation between $k=100$ and $k=1$ with $\kap=1$ in  the LPA for the ensuing exemplary initial conditions at $\Lambda=100$ in the proximity of the UV non-Gaussian fixed point in this truncation: $\mu_\Lambda=-0.2041 \Lambda^2$ (left panel) and $\mu_\Lambda=-0.2153 \Lambda^2$ (center panel) each with $\lambda_{2,\Lambda}=0.0444\Lambda^{-1}$ and $\lambda_{3,\Lambda}(\Lambda)=0.0008\Lambda^{-4}$. The main qualitative result is that the first potential displays symmetry restoration towards the IR whereas for the second the global symmetry remains broken. Right panel: Flow of the modulus of $\mu_k$ in the $\nmax=3$ truncation for: 
(I) the system of non-autonomous $\beta$-functions~\eqref{eq:betaCMcompact} with $\kap=1$ (dashed lines) and 
(II) the set of autonomous $\beta$-functions in the large-$\ak$ limit, Sec.~\ref{sec:UV/large volume} (continuous lines). 
Initial conditions are the same as in the left panel. The sign change of $\mu$ or the absence of it indicates the presence of a potential with broken symmetry or one without.}\label{figure:dimflowpotrank3dim3zeta1}
\end{figure}

The two phases are connected via a continuous phase transition at a critical surface (codimension-one subspace of the phase space).
This result applies for any rank $r$ as well as $\zeta=1/2,~1$ as long as $d>2$. 
Indeed, such a behaviour is well-known for $\text{O}(N)$-models with $N>1$ on flat space with $d>2$~\cite{Berges:2002ga,Codello:2012ec,Codello:2014yfa}. We illustrate this here for the case of the complex-valued rank-$3$ theory with $\zeta=1$ and $d=3$. In Fig.~\ref{figure:dimflowpotrank3dim3zeta1} we illustrate this situation by presenting the flow of the dimensionful potential and the coupling $\m$ in the $n=3$ truncation based on Eqs.~(\ref{eq:dimfulcouplingsuptoorder31}), (\ref{eq:dimfulcouplingsuptoorder32}) and (\ref{eq:dimfulcouplingsuptoorder33}). 
The latter is also contrasted with the flow of $\m$ in both phases in the large-volume limit with identical exemplary initial conditions.

We emphasize that, for this behaviour to occur, the combinatorially non-local structure of the interactions is not relevant, as the theory becomes effectively local, and is then simply traced back to the non-compactness of the product domain of the field. In fact, these results clearly show that phase transitions of this type can only arise in the tensorial theory if the domain is non-compact. Analogous results hold for the variants of models investigated in Sections~\ref{sec:tensors} and~\ref{sec:UV/large volume}. 
This finding is of particular relevance for the TGFT condensate cosmology approach~\cite{Gielen:2016dss,Oriti:2016acw,Pithis:2019tvp,Oriti:2021oux} which purports that condensate states and their collective dynamics lend themselves to model cosmological spacetimes. Correspondingly, this outcome obtained via FRG analysis also provides further evidence for the existence of
a physically desirable continuum limit in TGFT quantum gravity and in this way complement those gathered using Landau-Ginzburg theory of phase transitions~\cite{Marchetti:2021wp,Marchetti:2022igl,Marchetti:2022nrf}.

\section{Conclusion and discussion}

In this work, we explored the phase structure of rank-$\rk$ tensor fields on $\mathbb{R}^d$ with cyclic-melonic interactions using the FRG method. 
More precisely, we defined a field theory over $\mathbb{R}^d \times G^\rk$ wherein the fields transform as rank-$\rk$ covariant tensors. 
While the given type of interactions are local with respect to the $\mathbb{R}^d$-valued field arguments, they are combinatorially non-local in $G^\rk$ and feature invariance with respect to the tensorial transformation property. 
For the general setting where local and non-local degrees of freedom are propagating, we computed the FRG equation for $G=\U(1)$ in a local-potential approximation (LPA) at any order and at any RG scale $k$ which is in general non-autonomous. Based on this, we studied three different scenarios with their main results:  

$(1)$ In the first case we froze the non-local degrees of freedom which are thus not propagating. 
The theory is then a $\text{O}(N)^r$-invariant local field theory on $\mathbb{R}^d$ the FRG equation of which we examined in the large-$N$ limit.  There it becomes autonomous and assumes the form of the FRG equation of $\text{O}(N)$-invariant local field theory at large $N$, modified by an extra $\rk$ multiplicity of the quadratic mass term in the potential. This modification reflects the impact of the tensoriality of model. We analyzed its fixed point solutions and found results in qualitative agreement with those known for $\O(N)$ models~\cite{Berges:2002ga,Codello:2012ec,Codello:2014yfa}. In particular, one finds a Wilson-Fisher type fixed point for $2<d<4$, the precise numerical properties of which are only mildly affected by the factor $\rk$.

$(2)$ In the second scenario, local and non-local degrees of freedom are dynamical and we investigated the large-$\tilde{k}$ asymptotics of the corresponding FRG equation.
This yields another autonomous system of the same form as for the case $(1)$. However, the relevant parameters are $r$ and $\efd = \ld + (\rk -1)/\zeta$ therein such that one finds a Wilson-Fisher type fixed point only for $2<\efd<4$.

$(3)$ Finally, in the third case we studied the full non-autonomous FRG equation when local and tensorial degrees of freedom are propagating. 
This is in particular relevant for the TGFT approach. 
Specifically, we discussed the concept of effective dimension $\efd(k)$ introduced in~\cite{Pithis:2020sxm,Pithis:2020kio} which flows here from $\efd(k\to\infty)=\ld+(\rk-1)/\zeta$ in the UV to $\efd(k\to 0) = \ld$ in the IR. This allowed us to interpret the non-autonomous system to be similar to a standard autonomous one, however, for which the dimension $\efd$ is continuously interpolating between these asymptotic values. In the small-$k$ regime we found that the flow equation reduces to that of a local complex-valued scalar field theory on $d$-dimensional Euclidean
space. Consequently, the tensorial degrees of freedom of the model effectively vanish due to the isolated zero modes in the spectrum on a compact space. Based on this, we established that there are continuous transitions between phases of such a hybrid model with local and tensorial degrees of freedom for which global $\text{U}(1)$-symmetry is broken and unbroken as long as $d>2$ in accordance with the Mermin-Wagner-Hohenberg theorem~\cite{Mermin:1966fe,Hohenberg:1967zz,Coleman:1973ci}. We expect the same to directly apply to models with any tensor-invariant interactions living on the configuration space $\mathbb{R}^d \times G^\rk$ with compact $G$.

\

In the following we would like to briefly comment on limitations of our work and potential future extensions. 
An important restriction of our analysis is that we only considered the LPA. It would thus be desirable to go beyond this and work within the LPA$'$, thus incorporating the impact the wave-function renormalization into the flow equations. As pointed out in the main body of this text, we have indications that in order to find consistent flow equations two wave-function renormalizations have to be introduced reflecting the impact of fluctuations stemming from the local \textit{and} tensorial degrees of freedom. On the conceptual side, the elaboration of this case will further clarify the notion of scale for such hybrid theories of mixed degrees of freedom. This extension also goes in hand with clarifying the role of $\kappa$ in the kinetic operator and the examination of its flow.  

Another important extension of our work could be to go beyond the projection onto uniform field configurations which lies at the heart of this work and then study the impact of disconnected interactions, other melonic interactions and
eventually non-melonic interactions, see for instance~\cite{Carrozza:2017vkz,BenGeloun:2018ekd}. In light of such more involved investigations, our approximation used here can thus be seen as a first step towards realizing the phase diagram of the full
tensorial field theory.

Finally, from the point of view of TGFT, it is clear that for a realistic model of quantum geometry and gravity it is inevitable to consider $G$ to be the Lorentz group which encodes the causal structure of spacetime. It has already recently been shown for TGFTs on $\mathrm{SL}(2,\mathbb{C})$ with closure and simplicity constraints (to yield lattice gravity amplitudes for first-order Lorentzian Palatini gravity) that Landau-Ginzburg mean-field theory satisfactorialy describes a phase transition to a non-perturbative vacuum state making a compelling case for an interesting continuum geometric approximation~\cite{Marchetti:2022igl,Marchetti:2022nrf}. In light of this, we see our work also as a first step to study the phase structure and in particular the continuum limits of full-fledged GFT models for Lorentzian quantum gravity using the FRG.

\subsection*{Acknowledgments}
The authors thank D. Benedetti, K. Falls, R. Ferrero and L. Marchetti for discussions. A. Pithis acknowledges funding from DFG (German Research Foundation) research grants OR432/3-1 and OR432/4-1 and thanks for the generous financial support by the MCQST via the seed funding Aost 862983-4 granted by the DFG under Germany’s Excellence Strategy – EXC-2111 – 390814868. The work of J. Th\"{u}rigen was funded by DFG in two ways,
primarily under the author's project number 418838388 and
furthermore under Germany's Excellence Strategy EXC 2044--390685587, Mathematics M\"unster: Dynamics–Geometry–Structure.

\appendix

\section{Threshold Dirichlet integrals}\label{App:Dirichlet}
\renewcommand{\ks}{{2\zeta_0}}
\renewcommand{\ps}{{2\zeta_1}}
\newcommand{\qs}{{2\zeta_2}}

For fields with continuous spectra all necessary threshold functions can be read off from a general integral by  Gustav Lejeune Dirichlet \cite{Dirichlet:1839},
in original notation
\[
\underset{\underset{x>0,y>0,z>0, \dots}{\left(\frac{x}{\alpha}\right)^p +\left(\frac{y}{\beta}\right)^q +\left(\frac{z}{\gamma}\right)^r+...<1}
}{\int} \extd x \,\extd y \,\extd z\dots x^{a-1} y^{b-1} z^{c-1}\cdots
= \frac{\alpha^a}{p} \frac{\beta^b}{q} \frac{\gamma^c}{r}\cdots 
\frac{\Gamma(\frac{a}{p}) \Gamma(\frac{b}{q}) \Gamma(\frac{c}{r})\cdots}
{\Gamma(1+\frac{a}{p} + \frac{b}{q} +\frac{c}{r} + \dots )} 
\]
for all $a,b,c,...,p,q,r,...,\alpha, \beta,\gamma,...$ positive real numbers.
Now apply this to the threshold integral of a field with different kinetics given by positive real numbers $\zeta, \zeta_1$ and $\zeta_2$ and a coupling $\kap\in\R$ for momenta $\vp=(p_1,...,p_{d_1})$ and $\vq=(q_1,...,q_{d_2})$,
\bea
\underset{ |\vp|^\ps+\kap|\vq|^\qs < k^\ks
}{\int\extd\vp\,\extd\vq} 
&& \prod_{i=1}^{d_1}|p_i|^{x_i} \prod_{j=1}^{d_2} |q_j|^{y_j}
= \underset{ 
(\frac{|\vp|}{k^{\ks/\ps}})^\ps+ (\frac{|\vq|}{(k^\ks/\kap)^{1/\qs} })^\qs < 1
}{\int\extd\vp\,\extd\vq} \prod_{i=1}^{d_1}|p_i|^{x_i} \prod_{j=1}^{d_2} |q_j|^{y_j}
\crcr
&&= 2^{d_1+d_2}\frac{
\prod_i\frac{1}{\ps} \Gamma(\frac{x_i+1}{\ps})
\prod_j\frac{1}{\qs} \Gamma(\frac{y_j+1}{\qs})}
{\Gamma(1+\sum_i\frac{x_i+1}{\ps} + \sum_j\frac{y_j+1}{\qs} )}
( k^\ks )^{\sum_i\frac{x_i+1}{\qs}}
( k^\ks /\kap )^{\sum_j\frac{y_j+1}{\qs}} \, .
\crcr
&&
\eea
where the sums  $\sum_i$ and $\sum_j$, and products $\prod_i$ and $\prod_j$ are performed over $i=1,\dots, d_1$ and $j=1,\dots, d_2$, respectively.  

In particular, for exponents $x_i=y_j=0$ we have
\[\label{eq:generalized volume}
\underset{ |\vp|^\ps+\kap|\vq|^\qs < k^\ks
}{\int\extd\vp\, \extd\vq}
= \vol{d_1,d_2}^{(\zeta_1,\zeta_2)} (k^\ks )^{\frac{d_1}{\ps}}
( k^\ks/\kap )^{\frac{d_2}\qs}
\]
with generalized volume (expressing $z\Gamma(z)=\Gamma(1+z)$)
\[\label{eq:threshold function asymptotics 1}
\vol{d_1,d_2}^{(\zeta_1,\zeta_2)} := 2^{d_1+d_2} \frac{
\Gamma(1+\frac{1}{\ps})^{d_1}
\Gamma(1+\frac{1}{\qs})^{d_2}}
{\Gamma(1+\frac{d_1}{\ps} + \frac{d_2}{\qs} )}\, . 
\]
Furthermore for a single non-vanishing exponent $x_{i}=\ps$
\[\label{eq:threshold function asymptotics 2}
\underset{ |\vp|^\ps+\kap|\vq|^\qs < k^\ks
}{\int\extd\vp\, \extd\vq} |p_i|^\ps
=
\frac{1}{\ps} \frac1{1+\frac{d_1}{\ps} + \frac{d_2}{\qs}}
\vol{d_1,d_2}^{(\zeta_1,\zeta_2)}
( k^\ks )^{\frac{d_1}{\ps}+1}
( k^\ks /\kap )^{\frac{d_2}\qs}
\]
and, similarly, for a single $y_{j}=\qs$.
This generalizes the case of a single kind of kinetics $\ps=\qs=\ks$ which we have already considered before, Eq.~(A.23) in \cite{Pithis:2020sxm,Pithis:2020kio}.

Note that the relevant parameters are eventually only the ratios $d_1/\ps$ and $d_2/\qs$. 
Also, the exponent $\ks$ of the scale $k$ is irrelevant in the sense of rescaling since it is only the mass scale $\mu\sim k^{\ks}$ that occurs in the formula and which is physically relevant.

\renewcommand{\ks}{{2}}
\renewcommand{\ps}{{2}}
\renewcommand{\xi}{\alpha}

\section{Threshold functions of combined sums and integrals}
\label{app:threshold}

The threshold functions become technically much more involved when the spectra are discrete. 
In this work, we consider tensorial degrees of freedom on $G=\U(1)$.
Consequently, the relevant threshold function for the RG equations, \eqref{eq:thresholdfunction}, are
\begin{eqnarray}
\label{eq:thresholdfunctions}
\kf{\ld,s}_{p^{2\xi}}(k) &=&
\int_{\R^{\ld}} \extd\vp \; 
p^{2\xi} \sum_{\vrep \in (\mathbb{Z}\setminus\{0\})^{s} }
 \; 
\theta\left(k^\ks - p^\ps - \kap \cas\vrep \right) 
\;, \quad\quad \xi = 0, 1 \\
\kj{\ld,s}(k) &=&
\int_{\R^{\ld}} \extd\vp \sum_{\vrep \in (\mathbb{Z}\setminus\{0\})^{s} }
\sum_{c=1}^{s} |\rep_c|^\js
 \;  \theta\left(k^\ks - p^2 - \kap \cas\vrep \right)
\end{eqnarray}
for $\ld,s\in\mathbb{N}$ and the case $s=0$ is to be understood as
\begin{eqnarray}\label{eq:thresholdfunctions s=0}
\kf{\ld,0}_{p^{2\xi}}(k) 
&=& \int_{\R^{\ld}} \extd\vp \; 
p^{2\xi} \; \theta \left(k^\ks - p^\ps\right)  = v_{d} 
 \frac{1}{d+2\xi} (k^\ks)^{d+\xi} \\
  \crcr
\kj{\ld,0}(\ak) &=& 0 
\end{eqnarray}
For $s>0$, these are monotonically increasing but not continuous functions in the variable $k$. 
They are monotonic because the integrand is a positive monotonic function in all the cases. 
Nevertheless, they are not continuous since by definition each of the iterated summations of $\rep_c$ is up to a maximal integer $|\rep_c|=\lfloor k^\ks-p^2 - \kap\sum_{b<c}|\rep_b|^\js \rfloor$.
Thus, the threshold functions are in general intricate step functions.
But this should be regarded as an unphysical artifact of the application of a continuous coarse graining (the FRG method) to discrete momenta.
It is thus reasonable to approximate floor functions directly by their arguments.

\

\noindent
{\bf Threshold functions at $\zeta=1/2$.}
Deriving explicit expressions for the spectral sums is possible only for $\js=1$.
In this case there is a closed expression for the sum up to an integer $N\in\mathbb{N}$,
\[
\sum_{\substack{\vrep \in (\mathbb{Z}\setminus\{0\})^{s} \\ |\rep_1|+...+|\rep_s|\le N}} 1
= \sum_{\rep=1}^N \sum_{|\rep_1|+...+|\rep_s|=\rep} 1
= 2^s \sum_{\rep=1}^N \binom{\rep-1}{s-1}
= {2^s \binom{N}{s}} \, .
\]
Similarly,
\[\label{eq:sum of j}
\sum_{\substack{\vrep \in (\mathbb{Z}\setminus\{0\})^{s} \\ |\rep_1|+...+|\rep_s|\le N}} \sum_{c=1}^s |\rep_c|
= \sum_{\rep=1}^N \rep \sum_{|\rep_1|+...+|\rep_s|=\rep} 1
= 2^s \sum_{\rep=1}^N j \binom{\rep-1}{s-1}
= {2^s s \binom{N+1}{s+1}} \, .
\]
This applies to the spectral sums with $N=\lfloor \frac{k^\ks-p^\ps}{\kap}\rfloor$.
As argued, we are allowed to remove the floor function in a physical approximation.

One can expand the binomial function in positive real arguments $x\in\R$ as
\[\label{eq:expansion binomial}
\binom{x}{s} = \frac{1}{s!}\sum_{l=1}^s \aa_{s,l} x^l \, 
\]
with integer coefficients $\aa_{s,l}\in \Z$.
Then, the threshold function $\kf{\ld,s}_{p^{2\xi}}$, $s>0$, has the expansion
\[
\kf{\ld,s}_{p^{2\xi}}(k) 
= \int_{\R^{\ld}} \extd\vp \; p^{2\xi} \; 
\theta(k^\ks - p^2) 2^s \binom{\lfloor \frac{k^\ks-p^\ps}{\kap}\rfloor}{s}
= \frac{2^s}{s!} \sum_{l=1}^s \aa_{s,l} \ld \vol{\ld} \int_0^k \extd p\; p^{\ld+2\xi-1} \bigg\lfloor \frac{k^\ks-p^\ps}{\kap}\bigg\rfloor^l \, ,
\]
with $\vol\ld=\pi^{\ld/2}/\Gamma(\ld/2 +1)$.
Now, we approximate $\lfloor \frac{k^\ks-p^\ps}{\kap}\rfloor \approx \frac{k^\ks-p^\ps}{\kap}$ and use the binomial expansion to obtain
\begin{align}\label{eq:threshold function simplex}
\kf{\ld,s}_{p^{2\xi}}(k) &\approx 
\ld \vol\ld \frac{2^s}{s!} \sum_{l=1}^s \frac{\aa_{s,l}}{\kap^l} \sum_{i=0}^l (-1)^i \binom{l}{i} k^{\ks(l-i)} \frac{k^{\ks(\ld/2+\xi+i)}}{\ld+2\xi+2i} \nonumber\\
&= \frac{\ld\vol\ld}{\ld+2\xi}  
k^{\ks(\ld/2 +\xi)} \frac{2^s}{s!} \sum_{l=1}^s  \frac{\aa_{s,l}}{\binom{\frac{\ld}{2}+\xi+l}{l}}
\left(\frac{k^\ks}{\kap}\right)^{l} \, .
\end{align}
This result is a polynomial in $ 
k^\ks/\kap$ of degree $s$, times an overall power $k^{\ld+\ks\xi}$.
In particular, its large-$k$ asymptotics given by the leading order of the polynomial are in agreement with \eqref{eq:threshold function asymptotics 1} and \eqref{eq:threshold function asymptotics 2}.
Note that in this case the lowest order contribution is $l=1$.

A similar calculation is possible for $\kj{\ld,s}$ with $\js=1$. 
Expanding the sum \eqref{eq:sum of j}
\[\label{eq:expansion binomial +1}
\binom{x+1}{s+1} = \frac{1}{(s+1)!}\sum_{l=1}^{s+1} \bb_{s,l} x^l \, 
\]
with coefficients $\bb_{s,l}\in\Z$ we can expand also the threshold function $\kj{\ld,s}$ using the same approximations as before as 
\[
\kf{\ld,s}_{|j|} \approx 
  \vol\ld k^\ld  \,  \frac{2^s s}{(s+1)!} \sum_{l=1}^{s+1} \bb_{s,l} \sum_{i=0}^l  \binom{l}{i} \frac{(-1)^i \ld}{\ld+2i}
\left(\frac{k^\ks}{\kap} +1\right)^{l-i} \left(\frac{k^\ks}{\kap}\right)^{i} \, .
\]
This is a polynomial in $k^\ks/\kap$ of degree $s+1$, times the overall power $k^{\ld}$, with asymptotics again in agreement with \eqref{eq:threshold function asymptotics 2}. 

\

When $\js\ne1$, there are no closed expressions for the sums over discrete variables $\rep_c\in\Z$.
Such maps
\[
\mathbb{N} \to \mathbb{N} \quad ,\quad
    N \mapsto \sum_{\substack{\vrep \in (\mathbb{Z}\setminus\{0\})^{s} \\ |\rep_1|^\js+...+|\rep_s|^\js \le N^\js}}
\]
are certainly well defined, they are simply counting the number of points (excluding zeros) in a ball in $\js$ norm.
There simply are no closed expressions (known) for arbitrary $\js\in\R_+$.
Still, we can expect the above calculation for $\js=1$ to work in general.

The questions is then whether the result of the summation map allows for a polynomial expansion in $N$, like \eqref{eq:expansion binomial}.
If this is this case it will always lead also to a polynomial expansion of the full spectral sum like \eqref{eq:threshold function simplex}.
In particular, for each monomial $p^\alpha$, the integral over local momenta $p$ will give a contribution $k^{\ks(\ld+\alpha)/2}$ such that the result is again a polynomial of degree $s$ in $(k^\ks/\kap)^{1/\js}$ times $k^{\ks(\ld/2+\xi)}$.
In the end, it is not even necessary that the result of the discrete sum has a polynomial expansion;
it is sufficient to have a polynomial expansion after the continuous approximation of the step function via
\[
N^\js = \lfloor \frac{k^\ks-p^\ps}{\kap}\rfloor \approx \frac{k^\ks-p^\ps}{\kap} \, .
\]
But this means that, physically, the details of the summation maps will be anyway washed out by the necessary continuous approximation to threshold function.
But this allows to assume a polynomial expansion of the integrand in any case, leading to the result of $k^{\zeta(\ld+2\xi)}$ times degree-$s$ polynomial.

For general $\js$, another question is then what is the smallest power of monomials in this polynomial. 
In the case $\js=1$ leading to binomials for example, there is no constant contribution but the expansion starts at $l=1$, \eqref{eq:threshold function simplex}.
In general we cannot say anything about this issue.
Luckily, it is irrelevant for the physical properties of the renormalization group flow since the relevant scale-dependent functions $\kw{\rk-1}$, \eqref{eq:fullspectralsum}, sum over spectral sums from $s=0$ to $s=\rk-1$;
thus, the trivial case \eqref{eq:thresholdfunctions s=0} is always included and therefore the lowest order of the full sum $\kw{\rk-1}$ is always $k^{\ld}$.
This is the reason why in the case of non-autonomous equations the flow of the effective dimension~$\efd$, \eqref{eq:effective dimension}, is always to the value $\efd(k\to0)=\ld$ at small scale $k$.

\bibliographystyle{jhep}
\bibliography{main.bib} 

\end{document}